\documentclass{ws-ijmpd}
\usepackage[super,compress]{cite}
\usepackage[usenames,dvipsnames]{color}
  \definecolor{darkblue}{RGB}{0,0,150}

\usepackage[caption=false]{subfig}
\usepackage{amstext}

\begin{document}
 
\markboth{A. D. D. Masa, E. S. de Oliveira and V. T. Zanchin }
{Regular black holes and other electrically charged compact objects}

\title{New regular black hole solutions and other electrically charged compact objects with a de Sitter core and a matter layer}

\author{Angel D. D. Masa\footnote{angel.masa@ufabc.edu.br} , 
Enesson S. de Oliveira\footnote{enesson.oliveira@ufabc.edu.br} , 
Vilson T. Zanchin\footnote{zanchin@ufabc.edu.br}}

\address{Centro de Ci\^{e}ncias Naturais e Humanas, 
Universidade Federal do ABC\\
Avenida dos Estados 5001, 09210-580 -- Santo Andr\'{e}, S\~{a}o Paulo, 
Brazil}

\maketitle

\begin{abstract}
 
The main objective of this work is the construction of regular black hole solutions in the context of the Einstein-Maxwell theory.
The strategy is to match an interior regular solution to an exterior 
electrovacuum solution. With this purpose, we first write explicitly 
the Einstein field equations for the interior regular region. We take
an electrically charged nonisotropic fluid, which presents spherical
symmetry and a de Sitter type equation of state, where the radial 
pressure $p_r$ is equal to the negative of energy density $\rho$,
$p_r=-\rho$. Then, two solutions for the Einstein equations are built, a regular interior solution for the region with matter
satisfying a de Sitter equation of state, and an external solution for the region outside the matter, that corresponds to the Reissner-Nordstr\"om metric. To complete the solution we apply the 
Darmois-Israel junction conditions with a timelike thin shell at the
matching surface. It is assumed that the matching surface is composed
by a thin shell of matter, i.e., a surface layer in the form a perfect fluid obeying a barotropic equation of state, $\mathcal{P}=\omega\sigma$, $P$ and $\sigma $ being the intrinsic pressure and energy density of the shell, respectively, and $\omega$ a constant parameter. We show that there are electrically charged regular black hole solutions and other compact objects for specific choices of $\omega$ and of the other parameters of the model. Some properties of the objects are investigated.
 
\end{abstract}


\ccode{PACS numbers: 04.70.Bw, 04.20.Jb, 04.40.Nr}

\section{Introduction }

It is well known in general relativity that the end of life of massive stars is the gravitational collapse with the formation of black holes, i.e., a space-time region with an event horizon hiding a singularity, where all collapsed energy is trapped in a region of zero volume. As known from the no-hair theorem, this kind of compact objects are described by only three relevant quantities, which are the mass, the angular momentum, and the electric charge of the black hole. This fact is interesting also from the astrophysical perspective because astrophysical objects in general have angular momentum, besides mass, and may carry some 
amount of electric charge. In fact, Zakharov et al.\cite{Zakharov2005} analyzed 
the possibility of observing the parameters of astrophysical black holes.
They propose that, if the distance between the observer and the black hole is known, 
it is possible to extract information such as the mass, inclination angle and 
angular momentum of the black hole by measuring the retro-lensing image shapes 
(or mirage shapes). Moreover, if these parameters are measured with the 
necessary accuracy, it is possible to obtain also the electric charge of 
the black hole from observation of the mirage size. The authors 
claim that due to the extraordinary angular resolution, the space radio 
telescope RADIOASTRON will accurately measure the parameters of supermassive 
black holes like the one in the center of our galaxy.
Under real conditions, matter can fall into black holes or be around astrophysical
black holes. In such a context, we mention the work by 
Bronnikov and Zaslavskii\cite{Bronnikov2011} which considered different types 
of state equations of the ambient matter for which it is possible to have static 
equilibrium of non-rotating neutral or charged black holes and the surrounding matter.

From the theoretical point of view, the singularity
theorems\cite{Penrose1965,HawkingPenrose1970,HawkingEllis1973,Penrose1978}
assure that, if certain conditions on the energy-momentum distribution are satisfied,
singularities are unavoidable.
Since the presence of a singularity brings problems to the theory, it is natural 
to search for solutions free of singularities. There are in the literature several attempts to get rid of the singularities
inside black holes. Each of these attempts follows a specific strategy, 
and deals with a specific black hole, in violating the conditions of
the theorems on singularities. The resulting solutions replace the singularity
by a regular region and are called regular black holes.
The pioneering works are those of Sakharov\cite{sakharov} and
Gliner\cite{gliner1966}, which indicate that the cosmic singularity can be 
avoided with a de Sitter-like fluid, i.e., a perfect fluid with an state equation of 
the form $ p = - \rho $ ($ p $ being the pressure and $ \rho $ the energy 
density of the fluid). 
Following this idea, Dymnikova {\it et al}\cite{dy92,dy96,dy00,dy01,dy03,dy04,dy05,dy10} 
built regular black holes by replacing the singularity by a de Sitter-like core. 
The solutions obtained have as inner matter, a non-isotropic fluid satisfying the
equation of state $ p_{r} = - \rho$, with $p_{r}$ being the radial pressure. 
The fluid is continuously glued to the outer region corresponding to 
the Schwarzschild spacetime. 
Regarding the special properties of the interior matter required to violate, 
at least, the strong energy condition, the presence of anisotropic pressures 
is also a common feature of many regular black hole solutions. It is believed 
that regular black holes require some kind of exotic ingredient, such as 
nonlinear electrodynamics\cite{bardeen1968,ab00,ab98}, see also Refs.~\citen{brcritic,br01}, 
some kind of exotic field\cite{br071,br072}, phantom matter\cite{br06,ainou2011},
or, then, modifications of the theory of gravity also come into 
play\cite{mat06,mat08,Rodrigues:2015ayd}.  
The basic idea of this strategy is to find a source with interesting physical properties, 
be it a field or a matter model, which violates the strong energy condition, 
so that a regular hole solution can be constructed (see, e.g., 
Ref.~\citen{ansoldi2008} for a review on this topic).

Many other works have been done in an attempt to find regular interior 
solutions corresponding to the known black hole spacetimes. 
In the presence of electric charge, it is possible to find isotropic 
solutions\cite{Lemos:2011vz} of regular black holes. This regular solution contains
a central core of de Sitter type, glued to the exterior Reissner-Nordstr\"om 
metric thanks to an electrically charged surface but carrying no energy 
in the frontier of the core. A version with energy on the border 
was considered in Ref.~\refcite{Japoes2012}. Also, solutions of regular 
black holes with electrically charged phantom-type fluid in the core 
bounded to the outside of Reissner-Nordstr\"om were found in Ref.~\refcite{Lemos:2016vz}, 
where phantom matter is characterized by a perfect fluid for which $ p <- \rho $.

A common strategy to construct regular black hole solutions 
consists in imposing specific boundary conditions in the matching of 
two distinct spacetimes. One condition is making a smooth junction 
through a boundary surface between a de Sitter core joined to an outer
Schwarzschild solution\cite{Mars1996,Magli1999,Conboy2005,Elizalde2002}, 
or to a Reissner-Nordstr\"om solution \cite{Lemos:2011vz}. 
Other condition is making a transition from an inner solution to an exterior 
solution through a thin shell, see Refs.~\citen{Frolov1989,Frolov1990,Balbinot1990,Barrabes1996} 
for uncharged cases, and Ref.~\refcite{Japoes2012} for a charged case.
The matching procedure with thin shells is called the Darmois-Israel
formalism\cite{Darmois1927,Israel1966},  where it is derived convenient junction 
conditions and motion of the shell type spacelike and timelike between the spacetimes, 
for lightlike shells\cite{Barrabes1991}. Here we follow this strategy.

The main motivation of this work is to study the effects of a thin shell in
the construction of regular black holes. 
The interior region is made of a de Sitter core of charged matter. 
This is joined to the exterior Reissner-Nordstr\"om
metric by means of a thin shell of uncharged matter. The different classes of 
solutions are analyzed in terms of the free parameters of the model: 
the de Sitter radius, the global mass and charge of the solution, and the equation of state
for the matter on the shell.

This paper is organized as follows.
In Sec.~\ref{Sec. 2}, the basic equations of the model are implemented through
the Einstein-Maxwell geometry, whereas the solution of interior and exterior
spacetimes and the matching condition is discussed.
Section~\ref{Sec.3} is devoted to analyze the junction conditions and to
define the matter content of the thin shell.
In Secs.~\ref{firstclass}
and \ref{secondclass} we study the regions where regular black holes and
other interesting solutions are found. 
The concluding remarks are presented in Sec.~\ref{Sec. 4}.
 
 In this work geometric unities such that
the gravitational constant $G$ and the where speed of light $c$ are 
set to unity are employed, $G=1=c$, and the metric signature is $+2$.

\section{Spherical static spacetimes: Equations and solutions}\label{Sec. 2}

\subsection{The metric, matter fields and the basic equations}

In this work the main interest is in obtaining regular black hole solutions
for the Einstein-Maxwell equations with charged matter in the core. 
We assume the spacetime is static and spherically symmetric.
In this case, the line element is
\begin{equation}\label{eq:ds2}
ds^{2}=-B\left(r\right)dt^{2}+A\left(r\right)dr^{2}+r^{2}\left(d\theta^{2}
+\sin ^{2}\theta d\varphi^{2}\right),
\end{equation}
where $\{t,r,\theta, \varphi \}$
are the usual Schwarzschild coordinates and the potentials
$B\left(r\right)$ and $A\left(r\right)$  are functions that depend just on the radial 
coordinate $r$.
These potentials must satisfy the Einstein field equations  
\begin{equation}\label{eq:Einstein}
G_{\mu\nu}=8\pi T_{\mu\nu},
\end{equation}
where $G_{\mu\nu}$ is the Einstein tensor and the energy-momentum
tensor $T_{\mu\nu}$ has two contributions, $T_{\mu\nu}=M_{\mu\nu}+E_{\mu\nu}$. 
The first term is the energy-momentum tensor a non-isotropic fluid,
\begin{equation}
M_{\mu\nu}=\left(\rho_m+p_{t}\right)U_{\mu}U_{\nu} 
+\left(p_{r}-p_{t}\right)X_{\mu}X_{\nu}+p_{t}g_{\mu\nu},
\end{equation}
where $\rho_m$ is the energy density and $p_{r}$ and $p_{t}$
are the anisotropic (radial) and isotropic (tangential) pressure, respectively.
Additionally, $U_{\mu}$ is the four-velocity of the fluid
and $X_{\mu}$ is a spacelike unit vector which along the radial direction. 
Therefore, these vectors satisfy the normalization conditions, 
$U_{\mu}U^{\mu}=-X_{\mu}X^{\mu}=-1$, and  $U_{\mu}X^{\mu}=0$.
The second part of $T_{\mu \nu}$ is the electromagnetic energy-momentum tensor 
written in terms of Faraday-Maxwell tensor $F_{\mu\nu}$ by
\begin{equation}
E_{\mu\nu}=\frac{1}{4\pi}\left(F_{\mu}^{\gamma}F_{\nu\gamma}
-\frac{1}{4}g_{\mu\nu}F_{\gamma\beta}F^{\gamma\beta}\right),
\end{equation}
with  $F_{\mu\nu}$ obeying the Maxwell equations,
\begin{equation}\label{eq:Maxwell}
\nabla_{\nu}F^{\mu\nu}=4\pi J^{\mu}.
\end{equation}
In terms of the gauge potential $A_{\mu}$, the Faraday-Maxwell tensor reads $F_{\mu\nu}=\nabla_{\mu}A_{\nu}-\nabla_{\nu}A_{\mu}$,  where $\nabla_{\nu}$ is the covariant derivative. For the charged fluid, the  current density is given, in terms of the four-velocity $U^{\mu}$ and of the electric charge density 
$\rho_{e}$, by $J^{\mu}=\rho_{e}U^{\mu}$.

Once a static and spherically symmetric spacetime has been adopted \eqref{eq:ds2},
the electromagnetic potential $A_{\mu}$ takes the form\cite{eddington2} 
\begin{equation}
A_{\mu}=-\phi\left(r\right)\delta_{\mu}^{t},
\end{equation}
where $\phi\left(r\right)$ is the electric potential and $\delta$ stands
for the Kronecker delta. 
 Using the normalization conditions it is possible to write the 
 four-vectors $U_{\mu}$ and $X_{\mu}$ as 
\begin{equation}
U_{\mu}=-\sqrt{B\left(r\right)}\delta_{\mu}^{t},\qquad
X_{\mu}=\sqrt{A\left(r\right)}\delta_{\mu}^{r}.
\end{equation}

The Einstein-Maxwell equations \eqref{eq:Einstein} and 
\eqref{eq:Maxwell} yield a set of differential equations. 
The combination of the $tt$ and $rr$  components of these equations provides 
two important relationships
\begin{eqnarray}
\frac{B^{\prime}(r)}{B(r)}+\frac{A^{\prime}(r)}{A(r)}=8\pi 
rA(r)[\rho\left(r\right)+p_{r}\left(r\right)]\label{eq:CombinacaoEin}, \\ 
\left(\frac{r}{A(r)}\right)^{\prime}=1-8\pi 
r^{2}\left(\rho\left(r\right)+\frac{Q^{2}(r)}{8\pi r^{4}}\right),
\label{eq:CombinacaoEin2}
\end{eqnarray}
where (${}^{\prime}$) denotes the derivative with respect to the radial 
coordinate $r$, and $Q(r)$ is the electric charge inside a surface of
radius $r$. This charge $Q(r)$ is obtained from
the Maxwell equations \eqref{eq:Maxwell}, the only nontrivial equation,
and can be written as 
\begin{equation}\label{eq:carga}
Q\left(r\right)=4\pi\int_{0}^{r}\rho_{e}\bar{r}^{2}\sqrt{A\left(\bar{r}\right)
}d\bar{r}=\frac{r^{2}\phi^{\prime}\left(r\right)}{\sqrt{
A\left(r\right)B\left(r\right)}}.
\end{equation}

The mass $\mathcal{M}(r)$ inside a sphere of radius $r$ is defined by
\begin{equation}\label{eq:Massadentro}
\mathcal{M}(r)=4\pi\int_{0}^{r}\left( \rho_{m}(\bar{r})+\dfrac{Q^{2}(\bar{r})}{8\pi \bar{r}^{4}}\right)\bar{r}^{2}d\bar{r}+\dfrac{Q^{2}(r)}{2r}.
\end{equation}

A third independent equation comes from of the conservation law $\nabla_{\mu}T^{\mu\nu}=0$, 
which gives 
\begin{equation}\label{eq:equiHidro}
2p_{r}^{\prime}(r)+\frac{B^{\prime}(r)}{B(r)}\left[\rho(r) + p_{r}(r) 
\right]+\frac{4}{r}\left[ p_{r}(r) - 
p_{t}(r)\right]-2\frac{\phi^{\prime}(r)\rho_{e}(r)}{\sqrt{B(r)}}=0.
\end{equation}

Considering that $Q\left(r\right)$ is given in terms of $\rho_{e}(r)$, it is possible
to restrict the problem in terms of the six unknown functions: $A(r)$, $B(r)$,
$\rho_{m}\left(r\right)$, $p_{r}\left(r\right)$, $p_{t}\left(r\right)$, 
and $\rho_{e}(r)$ linked by only four field equations, namely, 
Eqs.~\eqref{eq:CombinacaoEin}, \eqref{eq:CombinacaoEin2}, \eqref{eq:carga}, 
and \eqref{eq:equiHidro}. 
This set of equations constitutes the most general system of equations 
for the kind of systems we are dealing with, a static, spherically symmetric, 
and charged matter distribution in general relativity. 
Therefore, there are certain degrees of freedom in the choices of these parameters that
may be used to simplify the resolution of the problem.
We deal with this issue next.

\subsection{Further hypotheses and new solutions}

\subsubsection{Inner Solution}
\label{minterior}

In this subsection we make some choices to find regular solutions. 
The first is regarding the equation of state for the anisotropic 
pressure $p_{r}\left(r\right)=
f\left[\rho\left(r\right)\right]$, and it is made so that the energy 
momentum tensor violates some of the energy conditions.  
Perhaps the simplest choice is that the fluid must obey the following 
equation \cite{sakharov, gliner1966}
\begin{equation}\label{eq:energiaescura}
p_{r}\left(r\right)=-\rho_{m}\left(r\right)
\end{equation}
Additionally, since there are six unknown functions and only four field
equations relating them, as it has been said before, it is possible to 
make some additional hypotheses with respect to the effective
energy density $\rho_{m}\left(r\right)+Q^{2}(r)/8\pi r^{4}$. 
We then consider the following relationship\cite{coospercruz,florides83}
\begin{equation}\label{eq:energiatotal}
8\pi\rho_{m}\left(r\right)+\frac{Q^{2}\left(r\right)}{r^{4}}=\frac{3}{R^{2}}, 
\end{equation}
 where $R$ is a constant. By replacing Eq. \eqref{eq:energiatotal} into Eq. 
\eqref{eq:CombinacaoEin2}, it follows
\begin{equation}\label{eq:pmAex}
A(r)=\left(1-\frac{r^{2}}{R^{2}}\right)^{-1},
\end{equation}
and from Eqs.~\eqref{eq:pmAex}, \eqref{eq:energiaescura}, and \eqref{eq:CombinacaoEin} we get
\begin{equation}\label{eq:pmBex}
B(r)=\frac{1}{A(r)}=1-\frac{r^{2}}{R^{2}}.
\end{equation}
This solution holds in the region filled by the electrically charged fluid,
up to a limiting surface $\Sigma$, of radius $r=a$. 
We still have one degree of freedom. So, following Ref.~\refcite{Florides1977},
for the refinement of the model it is assumed that the electric charge 
density $\rho_{e}(r)$ is given by \cite{Florides1977} 
\begin{equation}\label{eq:dencarga}
\rho_{e}(r)=\rho_{e0}\left(\frac{r}{a}\right)^{n}
\left(1-\frac{r^{2}}{R^{2}} \right)^{1/2},
\end{equation}
where $n\geq0$ is a dimensionless parameter and $\rho_{e0}$ is a constant
which gives the charge density at the centre of the distribution.
Replacing Eq.~\eqref{eq:dencarga} 
into Eq.~\eqref{eq:carga}, we get
\begin{equation}\label{eq:cargain}
Q(r)=q\left(\frac{r}{a}\right)^{n+3}, 
\end{equation}
where
\begin{equation}
q=\frac{4\pi\rho_{e0}a^{3}}{(n+3)}>0
\end{equation}
is the total electric charge of the matter distribution. 
Therefore, in the region $r<a$, the energy density $\rho_{m}\left(r\right)$, 
the radial pressure $p_{r}(r)$ and the tangent pressure $p_{t}(r)$ are
respectively
\begin{eqnarray}
8\pi\rho_{m}\left(r\right)&=&\frac{3}{R^{2}}-\frac{q^{2}}{a^{4}}
\left(\frac{r}{a} \right)^{2(n+1)},\label{eq:denEnerIn}\\
8\pi p_{r}\left(r\right)&=&-\frac{3}{R^{2}}+\frac{q^{2}}{a^{4}}\left(\frac{r}{a
}\right)^{2(n+1)},\label{eq:prerain}\\
8\pi 
p_{t}\left(r\right)&=&-\frac{3}{R^{2}}-\frac{q^{2}}{a^{4}}\left(\frac{r}{a
}\right)^{2(n+1)}. \label{eq:pretanin}
\end{eqnarray}
Besides that, the mass $\mathcal{M}(r)$ is
\begin{equation}
\mathcal{M}(r)=\dfrac{r^{3}}{2R^{2}}+\dfrac{q^{2}}{2a}\left(\dfrac{r}{a
}\right)^{2n+5}.
\end{equation}

It is seen that
the solution is free of physical and geometrical singularities at the origin 
since the energy density and the pressures are finite, $p_{r}\left(0\right)=
p_{t}\left(0\right)=-\rho_{m}\left(0\right)=-3R^{-2}$,
and the metric potentials are also finite there, $B(0)=A(0)=1$. It can be 
shown that the solution is regular everywhere in the region $0\leq r\leq a$.
Moreover,  the energy density is positive in that region if $3R^{-2}\geq q^{2}a^{-4}$.

\subsubsection{Exterior solution}\label{mexterior}

In the region outside the electrically charged 
fluid distribution, the fluid quantities are all zero, i.e., $\rho_{m}(r)=0, \, p_{r}(r)=0,\,
p_{t}(r)=0$, and $\rho_{e}(r)=0$.  Then, the solution of the Einstein-Maxwell 
equations for the region $r>a$ is given by
\begin{eqnarray}
A(r)&=&\left(1-\frac{2m}{r}+\frac{q^{2}}{r^{2}}\right)^{-1}=\frac{1}{B(r)},\\ 
\mathcal{M}(r)&=& m = {\rm constant},\\
 Q(r)&=&q={\rm constant}.
\end{eqnarray}
This is the (exterior) Reissner-Nordstr\"om metric parametrized by
the total gravitational mass $m$ and the total electric charge $q$ 
of the fluid contained within the (interior) region with matter.

\section{The junction and the surface layer}
\label{Sec.3}

\subsection{The junction conditions and the surface layer content}

In the study of solutions for compact objects in general relativity, 
thin shells  and surface layers are useful tools
to match the interior to the exterior regions of the objects. 
In the present case, the junction between de Sitter and
Reissner-Nordstr\"om spacetimes is made by means of a static 
thin shell at position $r=a$. Let us call it $\Sigma$. 
Such a surface is considered to carry an uncharged perfect fluid
characterized by the energy density $\sigma$ and pressure $\mathcal {P}$.
From the Darmois-Israel \cite{Darmois1927,Israel1966} junction conditions we get,
\begin{eqnarray}
\sigma&=&\frac{1}{4\pi a}\left( \sqrt{1-\frac{a^{2}}{R^{2}}} -
\sqrt{1-\frac{2m}{a}+\frac{q^{2}}{a^{2}}}\right)\label{eq:eds}, \\
\mathcal{P}&=&\frac{1}{8\pi a}\left( \frac{1}{\sqrt {1-\frac{a^{2}}{R^2}}}
- \frac{1-\frac{3m}{a}+\frac{2q^{2}}{a^{2}}}{\sqrt{1-\frac
{2m}{a}+\frac{q^{2}}{a^{2}}}} \right) -\sigma. \label{eq:pds}
\end{eqnarray}

The last two relations fully determine the energy-momentum content of
matching surface $\Sigma$ (a thin shell) in terms of four parameters: 
$a$, $R$, $m$, and $q$.  
However, instead of fixing such parameters, considering the ulterior
physical interpretation of the solutions, it is more interesting to assume
an specific model for the fluid in the shell. 
For this, it is usually assumed that there is a one-to-one relation between
the energy density and the pressure in the shell 
such that $\mathcal{P}=\mathcal{P}(\sigma)$. 
Our interest here is to study cases given by a linear barotropic state 
equation that simplifies the model and describe a hypothetical 
substance satisfying the relation
\begin{equation}\label{eq:estado}
\mathcal{P}=\omega\sigma,
\end{equation}
where $\omega$ is a constant. This barotropic equation of state may
represent different fluid types. Such fluids that can be
divided into three cases: (i) a usual fluid, i.e., one that satisfies
the energy conditions, when $\omega\geq0$; (ii) a dark energy fluid
when $-1<\omega<0$; and (iii) a phantom matter fluid when $\omega<-1$. 
It is worth mentioning that a thin shell with 
negative pressure may also describe the surface tension, characteristic of 
a boundary between two different material media \cite{Schmidt:1984be}.
A particular case of a shell made of dust fluid ($\omega=0$), including the stability conditions, was
studied in Ref.~\refcite{Japoes2012}. An equation of state as in \eqref{eq:estado} was employed in the study of regularity and stability 
of a thick matter layer surrounding static black holes performed in Ref.~\refcite{Bronnikov2011}. However, as we see below, the situation is different for the regular black holes analyzed here because all the matter content, including the timelike thin shell, resides inside the Cauchy horizon.

Equation\eqref{eq:eds} may be readily integrated along the surface to
give the total mass of the shell
\begin{equation}\label{eq:M1}
\frac{M}{a}=\sqrt{1- \frac{a^2}{R^2}} - \sqrt{1 -\frac{2m}{a} 
+ \frac{q^2}{a^2}}. 
\end{equation}
with $M=4\pi a^{2}\sigma$.
Similarly, from Eqs.~\eqref{eq:estado} and\eqref{eq:pds} we obtain 
\begin{equation}\label{eq:M2}
2\left(1+\omega\right)\frac{M}{a} =
\frac{1}{\sqrt {1-\frac{a^{2}}{R^2}}} -
\frac{1-\frac{3m}{a}+\frac{2q^{2}}{a^{2}}}{\sqrt{1-\frac
{2m}{a}+\frac{q^{2}}{a^{2}}}}.
\end{equation}
These two relations may be used to express two of the fundamental parameters, 
$a$, $R$, $M$, $m$, $q$, and $\omega$, in terms of the other four free parameters.
There are, of course, a number of choices for the four free parameters, and
in any case a physical interpretation of the resulting solutions should be given.
In the following we discuss on this issue.

\subsection{The parameter space}
\label{sec:prelim}
 
 As commented above, the choice of an equation of state and together with the
 junction conditions on the shell set some constraints on the parameters of 
 the model. Equations~\eqref{eq:M1} and \eqref{eq:M2} furnish two  constraints, 
resulting in four free parameters only. Among the possible choices, 
we consider $\omega$, $a$, $q$, and $R$ as free parameters. 
Moreover, without loss of generality, all parameters with dimension of 
length may be normalized by $R$. Hence, in the numerical analysis we take
$a/R$, $q/R$ (or $q^2/R^2$), and $\omega$ as the only free parameters of the model.  

Now the ranges of parameters have also to be considered.
Since the matching surface $\Sigma$ is assumed to be a timelike or, in the
limiting case, a null (lightlike) surface, the radius $a$ 
cannot be larger than $R$,
so that we fix $a\in [0,R]$, or $a/R\in [0,1]$. This constraint
is necessary
to assure that the Killing vector $K^\mu = \delta^\mu_t$ is timelike (or null)
on  $\Sigma$, i.e., $-K_\mu K^\mu = -g_{tt}= 1-a^2/R^2\geq 0$.
The range of the electric charge is, in principle, the whole real line. 
However, since the electric charge appears always as $q^2$ in all
of the equations, without loss of generality we may take
$q/R\in [0,\infty)$, or $q^2/R^2\in [0,\infty)$. To avoid having 
a surface layer with pressure larger than the energy density we restrict
$\omega$ to the interval $\omega\in(-\infty, 1]$. Negative values of $\omega$
are allowed in order to describe dark energy like fluids.

The constraint $1- a^2/R^2 \geq 1$ also implies further constraints on the  
parameters of the exterior metric, $m$ and $q$. They must be such that 
condition $-g_{tt}=1- 2m/a +q^2/a^2 > 0 $ is satisfied.
Hence, in the undercharged case where $m^2 > q^2$, one must have
$a < r_-$, or $a > r_+$, where
\begin{equation}\label{eq:rpm}
    r_\pm = m \pm \sqrt{m^2 - q^2}
\end{equation}
are the zeros of the Reissner-Nordstr\"om metric coefficient 
$-g_{tt}=1- 2m/r +q^2/r^2$. In such a case, the matching is made
inside the Cauchy horizon $r_-$, corresponding to a regular black hole,
or outside $r_+$ (there are no horizon in this case), 
corresponding to a regular charged star. In the limiting extreme case where $m^2=q^2$, the matching may be at the Cauchy horizon, i.e., with $a=r_-$, corresponding to a quasiblack hole (see Ref.~\citen{Lemos:2011vz}).
On the other hand, in the overcharged case
where $m^2< q^2$ there are no further constraints on such parameters.

\subsection{A preliminary analysis of the solutions}

The next step to understand the equilibrium solutions is to
examine their physical properties. We start considering the mass of the shell.
For this, we use Eqs.~\eqref{eq:M1} and \eqref{eq:M2} to eliminate the 
mass parameter $m$ and solve for the shell's total mass $M$ 
and obtain two solutions,
\begin{multline}\label{eq:massadashell}
M_{\pm} = 
\dfrac{a}{\left(1+4\omega\right)\sqrt{1-\dfrac{a^2}{R^2}}}
\Bigg\{ 2\omega\left(1-\dfrac{a^2}{R^2}\right)
+\dfrac{a^2}{R^2} \\ 
\pm \sqrt{\left[2\omega\left(1-\dfrac{a^2}{R^2}\right)+\dfrac{a^2}{R^2}\right]^{2}
- \left(1+4\omega\right)\left(1-\dfrac{a^2}{R^2}\right)
\left(\frac{3a^2}{R^2}-\frac{q^2}{a^2}\right)}\Bigg\}.
\end{multline}

The total mass of the distribution, $m$, may also be written in terms
of the free parameters $a/R$, $q/R$, and $\omega$,
\begin{multline}\label{eq:totalmass}
m_\pm=\dfrac{a}{2} \left( 1+\dfrac{q^{2}}{a^{2}}-
\left[\sqrt{1-\dfrac{a^{2}}{R^{2}}}-\dfrac{1}{\left(1+4\omega\right)
\sqrt{1-\dfrac{a^2}{R^2}}}\Bigg\{ 2\omega\left(1-\dfrac{a^2}{R^2}\right)
+\dfrac{a^2}{R^2} \right. \right.\\ 
\left. \left. \pm \sqrt{\left[2\omega\left(1-\dfrac{a^2}{R^2}\right)
+\dfrac{a^2}{R^2}\right]^{2}
- \left(1+4\omega\right)\left(1-\dfrac{a^2}{R^2}\right)
\left(\frac{3a^2}{R^2}-\frac{q^2}{a^2}\right)}\Bigg\}\right]^{2}\right), 
\end{multline}
where the solution $m_\pm$ corresponds to $M_\pm$, respectively.

A critical point of Eqs.~\eqref{eq:massadashell} and \eqref{eq:totalmass} is $\omega=-1/4$, 
such that functions $M_\pm$ and $m_\pm$ seem to be singular at that point. As one can see by a detailed analysis, expressions for $M_\pm$ and $m_\pm$ are not well defined functions of $\omega$, since they have different limits when approaching $\omega=-1/4$ from above and from below, depending on the values of the other parameters.  
The expressions for $M_+$ and $m_+$ have no definite limits from both sides and diverge at that point. However, under the further constraint $1/3\leq a^2/R^2<1$, it can be shown that the functions $M_-$ and $m_-$ are well defined solutions even for $\omega=-1/4$. In fact,  this apparent problem may be circumvented by replacing $\omega=-1/4$ since the beginning into
Eq.~\eqref{eq:M2} and using Eq.~\eqref{eq:M1} to eliminate $m$. The resulting 
well defined shell mass is then
\begin{equation}
M_\pm={a\left(\dfrac{q^2}{a^2}-\dfrac{3a^2}{R^2}\right)}
\sqrt{1-\dfrac{a^{2}}{R^2}}{\left(1-\dfrac{3a^2}{R^2}\right)^{-1}}. 
\end{equation} 
This particular solution is regular everywhere except at $a^2/R^2=1/3$. 
In turn, for $\omega=-1/4$ the total mass $m_\pm$ results
\begin{equation}\label{eq:m_w1o4}
    m_\pm = \dfrac{a}{2} \left( 1+\dfrac{q^{2}}{a^{2}}- \left(1-\dfrac{a^2}{R^2}\right)
\left[1-{\left(\dfrac{q^2}{a^2}-\dfrac{3a^2}{R^2}\right)}
{\left(1-\dfrac{3a^2}{R^2}\right)^{-1}} \right]^{2}\right), 
\end{equation}
which diverges just when $a^2/R^2 =1/3$, with $q^2/a^2\neq 1$. Note also that there is only one solution, the one corresponding to the appropriate limit of the solution with minus signs in expressions \eqref{eq:massadashell} and \eqref{eq:totalmass}.

With the expressions for the mass functions in terms of the free parameters
of the model at hand, the analysis is straightforwardly performed. 
As we shall see next, the solution $M_-$ ($m_-$) is more interesting than $M_+$
($m_+$) since it comprises a larger variety of compact object with fair physical
properties. The main physical properties of the two solutions are discussed
in the following, first we take the more interesting case for 
$M_-$ and then we analyze the case for $M_+$.

\section{The first class of solutions: for $M=M_-$ and $m=m_-$}
\label{firstclass}

\subsection{General remarks}
\label{sec:generalM-}

Here we consider the main properties of the spacetimes spanned by taking 
the solution with minus signs in Eqs.~\eqref{eq:massadashell} and \eqref{eq:totalmass}. 
First note that such relations must be carefully analyzed 
 for $a/R =1$, where both expressions seem to be singular.   
 In fact, taking the limit $a/R\rightarrow 1$ we find $  M_-/R=0$,
which means that, in principle, the matching surface may be placed at $a/R=1$, 
as long as the mass of the shell vanishes. 
In such a limit, from Eq.~\eqref{eq:totalmass} one has that
$    m_- = q^2/2a + a/2 $
and so the energy density,  Eq.~\eqref{eq:eds}, results zero independently of 
the values of $q$ and $\omega$. However, it is seen from Eq~\eqref{eq:pds} that the intrinsic pressure of the shell
diverges in the limit $a/R\rightarrow 1$ and hence the respective
solutions are singular.

Some more details of the solutions for a few different values of the 
parameter $\omega$ are presented below.

\subsection{Analysis of the radii $r_{+}$ and $r_{-}$}\label{sec:radii-1}

In order to investigate the properties of the solutions in terms of
the free parameters, a key issue is to test for the presence or 
absence of horizons. For instance, for a given solution to represent a
regular black hole, the geometry necessarily has to present horizons. 
This means that the radii $r_\pm$ must assume real positive values, what guarantees that $m_{-}/R\geq q/R$.
 Moreover, and more important, at least $r_+$ must be
larger than the radius of the matter region boundary, i.e., $r_+/R > a/R$.
Furthermore, the imposition of a timelike boundary layer (shell) implies the 
matching of the de Sitter solution (inner metric, see Sec.~\ref{minterior}) and the 
Reissner-Nordstr\"om solution (exterior metric, see Sec.~\ref{mexterior}),
at the radius $r/R=a/R$, has to be located inside $r_-$ and, therefore, 
one has the constraint $a/R\leq r_{-}/R$.  
As a consequence, the Reissner-Nordstr\"om gravitational radius $r_+$ and the inner 
radius $r_-$ are both in the exterior of the region of matter distribution, been actually the event and Cauchy horizons, respectively.

\begin{figure*}[htb]
\centering
\subfloat[$\omega=0$; $q/R=0.650$]{\label{f:w=0q=0650}
\protect\protect\includegraphics[width=0.38\textwidth]{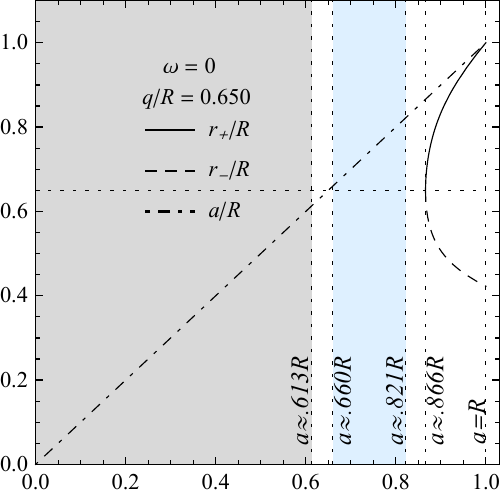}}\quad
\subfloat[$\omega=0$; $q/R=0.940$]{\label{f:w=0q=0940}
\protect\protect\includegraphics[width=0.38\textwidth]{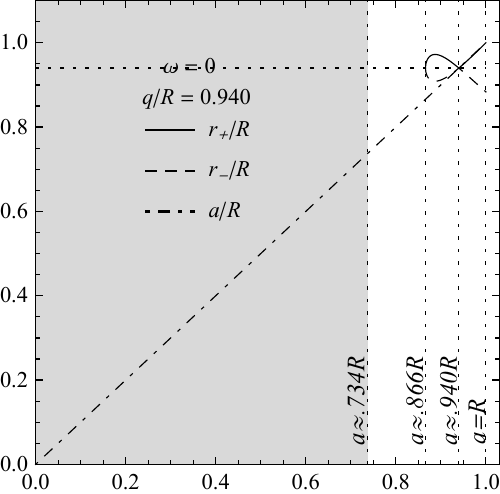}}\\
\subfloat[$\omega=0$; $q/R=1.000$]{\label{f:w=0q=1000}
\protect\protect\includegraphics[width=0.38\textwidth]{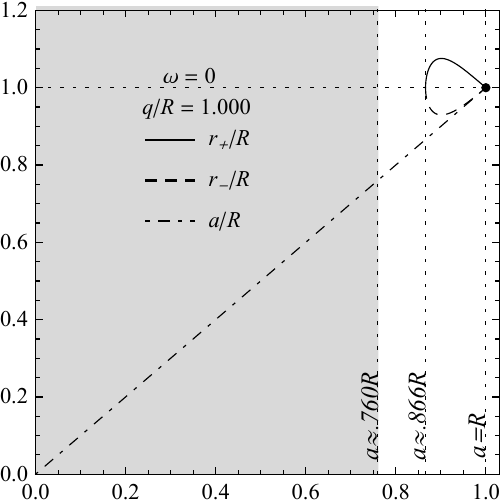}}\quad
\subfloat[$\omega=0$; $q/R=1.500$]{\label{f:w=0q=1500}
\protect\protect\includegraphics[width=0.38\textwidth]{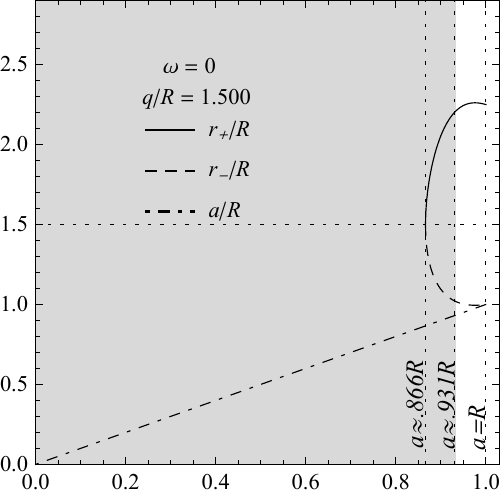}}
\protect\protect\caption{\small Plots of the radii $r_{-}/R$, $r_{+}/R$ and 
$a/R$ as a function of $a/R$ for $\omega=0$ and four different values of $q/R$.}
\label{f:w=0q}
\end{figure*}

Following Ref.~\refcite{Lemos:2016vz}, in this section 
we plot the radii $r_{+}/R$, $r_{-}/R$ [see Eq.~\eqref{eq:rpm}], and $a/R$ as 
a function of $a/R$ for some values of the electric charge $q/R$. 
We do this analysis for a set of representative values of $\omega$, namely, 
$\omega= 3/4,\, 0\,, -3/4,$ and $-2$. The results are shown respectively in 
Figs.~\ref{f:w=34q}, \ref{f:w=0q}, \ref{f:w=-34q}, and \ref{f:w=-2q}. 
The conventions adopted in such figures, based in the values of the shell
mass $M_-$, are the following. White regions refer to positive $M_-$, 
light blue (light grey in the black $\&$ white version) regions refers to
imaginary values of $M_-$, solutions that we do not consider here, and grey 
(dark grey in the black $\&$ white version) regions refer to negative values of 
the mass $M_-$.

The chosen values of $\omega$ are representative of the wide classes of solutions
that can be found in the present model.

\subsubsection{The case $\omega=3/4$: Standard matter}

\begin{figure*}[ht]
\centering
\subfloat[$\omega=3/4$; $q/R=0.800$]{\label{f:w=34q=0800}
\protect\protect\includegraphics[width=0.38\textwidth]{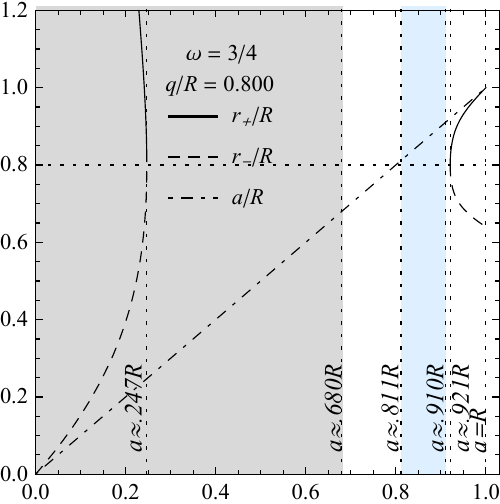}}\,
\subfloat[$\omega=3/4$; $q/R=0.950$]{\label{f:w=34q=0950}\protect
\protect\includegraphics[width=0.38\textwidth]{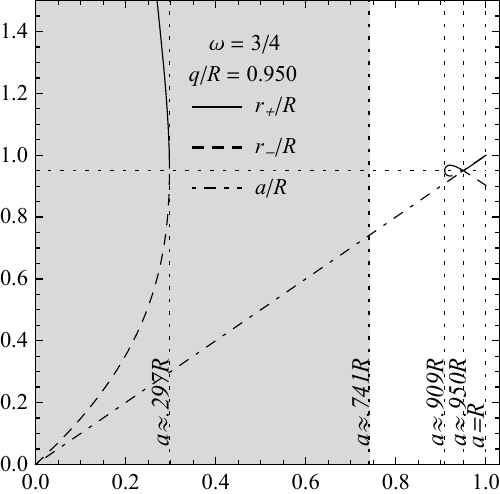}}\,
\subfloat[$\omega=3/4$; $q/R=1$]{\label{f:w=34q=1000}\protect
\protect\includegraphics[width=0.38\textwidth]{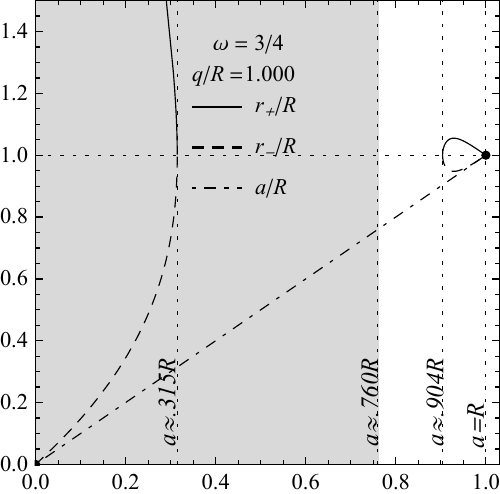}}\,
\subfloat[$\omega=3/4$; $q/R=1.400$]{\label{f:w=34q=1400}\protect
\protect\includegraphics[width=0.38\textwidth]{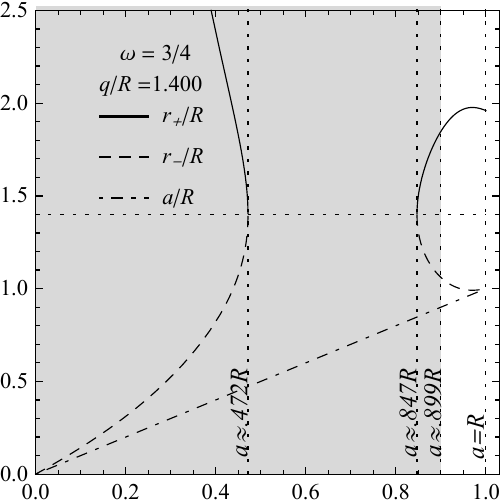}}\,
\protect\protect\caption{\small Plots of the radii $r_{-}/R$, 
$r_{+}/R$, and $a/R$ as a function of $a/R$ for $\omega=3/4$ and four different 
values of $q/R$.}
\label{f:w=34q}
\end{figure*}

The equation of state ${\mathcal P}= 3\sigma/4$ represents a fluid that satisfies
the standard energy conditions as long as $\sigma\geq 0$.
The graphs for this case are shown in Fig. \ref{f:w=34q}, where we plot the several
radii for four interesting values of $q/R$, namely, $q/R=0.80$, $q/R=0.95$,
$q/R=1.0$, and $q/R=1.4$, as indicated.
The dot-dashed line represents the radius of the matching shell $a/R$,
and the horizons radii $r_{-}/R$ are $r_{+}/R$ are represented by a dashed
line and a solid line, respectively. Vertical dashed lines indicate some particularly interesting values of $a/R$.

The top left panel of Fig.~\ref{f:w=34q}, Fig.~\ref{f:w=34q=0800}, is for $q/R=0.800$ besides $\omega=3/4$. In this case the graph shows six different regions, separated by vertical dashed lines.
The interval $0 < a/R \lesssim 0.247$ contains regular black holes with 
a thin shell of negative mass at the boundary inside the Cauchy horizon (at $a< r_-$).
At $a/R\sim 0.247$ one has $r_-/R=r_+/R$ and the solution is an extreme black hole.
All of the interval $0.247\lesssim a/R\lesssim 0.921$ contains overcharged stars for which $m_{-}/R<q/R$, but with different class of thin shells as mentioned above.
At $a/R\sim 0.921$ the solution is an extremely charged star.
In the interval $0.921\lesssim a/R< 1$ the solutions are regular 
undercharged stars for which  $m_{-}/R>q/R$ and with a thin shell with 
positive mass at the boundary. At $a/R=1$ the ratio $r_+/R$ is also unity and such a line
contains singular solutions because the intrinsic pressure of the shell diverges in that limit. As the radius of the boundary $a/R$ tends to unity, one also has $a/R\to r_+/R$ and the matching is made on the horizon of the Reissner-Nordstr\"om metric,
which is a lightlike surface. As discussed in Ref.~\citen{Lemos:2007yh},
 a possible interpretation for this solution is as a nonextreme (singular) quasiblack hole. It is a singular solution in the sense that the pressure of the boundary shell is infinitely large. 

 The top right panel of Fig.~\ref{f:w=34q}, Fig.~\ref{f:w=34q=0950}, 
is for $q/R=0.95$ besides $\omega=3/4$. In this case the graph shows five regions.
The interval $0 < a/R \lesssim 0.297$ contains regular black holes with 
a thin shell of negative mass inside the Cauchy horizon at $a/R< r_-/R$.
At $a/R\simeq 0.297$ one has $a/R<r_-/R=r_+/R$ and the solution is an extreme regular black hole.
All of the interval $0.297\lesssim a/R\lesssim 0.909$ contains overcharged stars for which $m_{-}/R<q/R$, 
but with different class of thin shells as mentioned above. No shells of imaginary 
mass are present in this case.
At $a/R\sim 0.909$ the solution is also an extreme regular black hole.
The solutions belonging to the interval $0.909\lesssim a/R< 0.950$ are regular 
charged black holes. The solution for $a/R=r_+/R=r_-/R = 0.950$ is extremal
in the sense that $m_{-}/R=q/R$, and the energy density and the intrinsic mass of the shell are both finite quantities.  
Additionally, a careful investigation of Eq.~\eqref{eq:pds} reveals that the intrinsic pressure is also well defined and assumes a finite value in the limit $a\to q =m_{-}$. Note that the matching is made on the extreme horizon of the Reissner-Nordstr\"om metric,
which is a lightlike surface. On the other hand, from the point of view of the inner de Sitter metric, the junction surface (at $a/R=.950$) is timelike. According to Ref.~\citen{Lemos:2007yh},
 this solution is a kind of extreme quasiblack hole, with a naked behavior at the extreme quasi-horizon $a=r_+=r_-$. 
The interval $0.950 < a/R < 1$ contains very compact regular undercharged stars, with
the matching surface radius very close to the gravitational radius $a/R\gtrsim r_+/R$.
At $a/R=1$ the ratio $r_+/R$ is also unity and, as in the case of Fig.~\ref{f:w=34q=0800}, the solution represents a singular nonextreme quasiblack hole.

 The bottom left panel of Fig.~\ref{f:w=34q}, Fig.~\ref{f:w=34q=1000},
is for $q/R=1.00$ besides $\omega=3/4$. 
In the interval $0< a/R< 1$ the solutions are identical to the case of 
Fig.~\ref{f:w=34q=0950} within the interval $0< a/R < 0.950$ and 
we do not repeat the description here. 
At $a/R=1$ it happens that $a/R=r_+/R=r_-/R=m_-/R=q/R$ and the matching surface coincides with an extreme (lightlike) horizon.
The intrinsic energy density and the mass of the shell are finite quantities (vanish), but 
the intrinsic pressure is not well define in this limit, it diverges. This kind of solutions may be interpreted as singular quasiblack holes\cite{Lemos:2007yh}.

The bottom right panel of Fig.~\ref{f:w=34q}, Fig.~\ref{f:w=34q=1400}, is for $q/R=1.40$ 
besides $\omega=3/4$. 
In this case the graph shows four different regions.
In the intervals $0< a/R\lesssim 0.472$ and $0.847\lesssim a/R\lesssim 0.899$
the solutions are regular black holes with a thin shell of negative mass
inside the Cauchy horizon. At $a/R \simeq 0.472$ and $a/R\simeq 0.847$ the solutions
are regular extreme black holes.
The interval $0.472\lesssim a/R\lesssim 0.847$ contains overcharged stars 
with a thin shell of negative mass at the boundary.
In the intervals $0.899< a/R <1$ the solutions are regular black holes with a thin shell of positive mass at the boundary inside the Cauchy horizon.
At $a/R =1$ it happens that $a/R=r_-/R$ the junction is made at the inner (Cauchy) horizon 
of the exterior metric, and so it is a black hole solution. It is a singular black hole since the intrinsic pressure of the shell is infinitely large pressure of the shell.

\subsubsection{The case  $\omega=0$: Dust}

With this equation of state, the intrinsic fluid on the thin shell 
has zero pressure and represents dust matter. 
A similar situation with $\sigma\neq 0$ and ${\mathcal P} =0$ but taking an isotropic
fluid inside the shell was study in Ref.~\refcite{Japoes2012}, 
see also Ref.~\refcite{Lemos:2011vz} for the particular case with
$\sigma =0$ and also ${\mathcal P}=0$. 
Here we generalize that analysis for the 
non-isotropic case, and show for instance that there are regular black 
holes for other ranges of the electric charge $q$, as it can be seen
in Fig. \ref{f:w=0q}. The kind of objects found in this case are similar to the case of Fig.~\ref{f:w=34q} and then we shorten the descriptions of the corresponding properties.

The top left panel of Fig.~\ref{f:w=0q}, Fig.~\ref{f:w=0q=0650}, is for $q/R=0.650$ 
besides $\omega=0$. 
In this case the graph shows five different regions.
The interval $0< a/R\lesssim 0.613 $ contains overcharged stars with a thin shell 
of negative intrinsic mass. 
The solution at $a/R\simeq 0.613 $ is an overcharged star with no thin shell, i.e., 
the mass of the shell, as well as the intrinsic energy density and pressure of the shell are all zero. 
The intervals $0.613\lesssim a/R\lesssim 0.660 $ and $0.821\lesssim a/R\lesssim 0.866 $ 
contains overcharged stars with a thin shell of positive intrinsic mass. 
The interval $0.660\lesssim a/R\lesssim 0.821$ contains no solutions, i.e., 
the mass of the shell assumes complex values. 
The solution at $a/R \simeq 0.866$ is an extreme regular black hole.
The interval  $0.866\lesssim a/R<1 $ contains regular black holes with
a thin shell of positive intrinsic mass.
The solution at $a/R=1$ is similar to the case of Fig.~\ref{f:w=34q=0800} and may be interpreted as a singular quasiblack hole.

The top right panel of Fig.~\ref{f:w=0q}, Fig.~\ref{f:w=0q=0940}, is for $q/R=0.940$ 
besides $\omega=0$. 
In this case the graph shows four different regions.
The interval $0< a/R\lesssim 0.734$ contains overcharged stars with a thin shell 
of negative intrinsic mass.
The solution at $a/R\simeq 0.734$ is an overcharged stars with no thin shell, i.e., 
the mass of the shell, as well as the intrinsic energy density and pressure of the shell are all zero.
The interval $0.734\lesssim a/R\lesssim 0.866$ contains overcharged stars with a 
thin shell of positive intrinsic mass.  
The solution at $a/R \simeq 0.866$ is an extreme regular black hole.
The interval  $0.866\lesssim a/R < 0.940 $ contains regular black holes.
The solution for $a/R=r_+/R=r_-/R= 0.94$ is extremal
in the sense that $m_{-}/R=q/R$ and the matching is done at an extreme
Killing horizon (a lightlike surface) for the exterior spacetime. Similarly to the case of Fig.~\ref{f:w=34q=0950}, the energy density, the intrinsic mass, and the pressure of the shell are all finite quantities in such a limit. The corresponding solution is an extreme quasiblack hole.  
The solution at $a/R=1$ is a singular quasiblack hole. The matching occurs at the horizon of the exterior metric, $a/R=r_+/R$. The energy density and the mass of the shell vanish but the pressure is not well defined in that limit.

The bottom left panel of Fig.~\ref{f:w=0q}, Fig.~\ref{f:w=0q=1000}, is for $q/R=1.000$ 
besides $\omega=0$. 
In this case the graph shows three different regions.
The interval $0<a/R\lesssim0.760$ contains overcharged stars with a thin shell 
of negative intrinsic mass. 
The solution at $a/R\simeq 0.760 $ is an overcharged star with no thin shell, i.e., the 
mass of the shell, as well as the intrinsic energy density and pressure are zero. 
The interval $0.760\lesssim a/R\lesssim 0.866$ contains overcharged stars with a 
thin shell of positive intrinsic mass.
The solution at $a/R \simeq 0.866$ is an extreme regular black hole.
The interval  $0.866\lesssim a/R<1 $ contains regular black holes.
The solution at $a/R=1$ also satisfy the relations $a/R=r_-/R=r_+/R=m_{-}/R=q/R=1$, the matching occurs at the extreme horizon of the exterior metric, and the 
pressure diverges in that limit. The solution is a singular quasiblack hole.

The bottom right panel of Fig.~\ref{f:w=0q}, Fig.~\ref{f:w=0q=1500}, 
is for $q/R=1.500$ besides $\omega=0$. 
In this case the graph shows three different regions.
The interval  $0< a/R\lesssim0.866$ contains overcharged stars with a thin shell
of negative intrinsic mass. 
The solution at $a/R\simeq 0.866 $ is an extreme regular black hole with a 
thin shell of negative intrinsic mass. 
The interval $0.866\leq a/R\lesssim 0.931$ contain regular black holes with 
a thin shell of negative intrinsic mass. 
The solution at $a/R \simeq 0.931$ is an regular black hole with no thin shell, i.e., the 
mass of the shell, as well as the intrinsic energy density and pressure are zero.
The interval  $0.931\lesssim a/R<1 $ contains regular black holes with a thin shell of positive mass.
The solution at $a/R=1$ is a singular lack hole. The matching occurs inside the event horizon, at the Cauchy horizon of the exterior metric, and the intrinsic pressure of the shell diverges in that limit.

\begin{figure*}[htb]
\centering
\subfloat[$\omega=-3/4$; $q/R=0.500$]{\label{f:w=-34q=0500}
\protect\protect\includegraphics[width=0.38\textwidth]{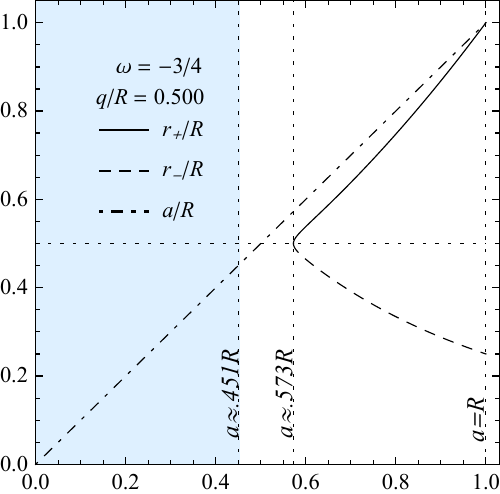}}\quad
\subfloat[$\omega=-3/4$; $q/R=0.900$]{\label{f:w=-34q=0900}
\protect\protect\includegraphics[width=0.38\textwidth]{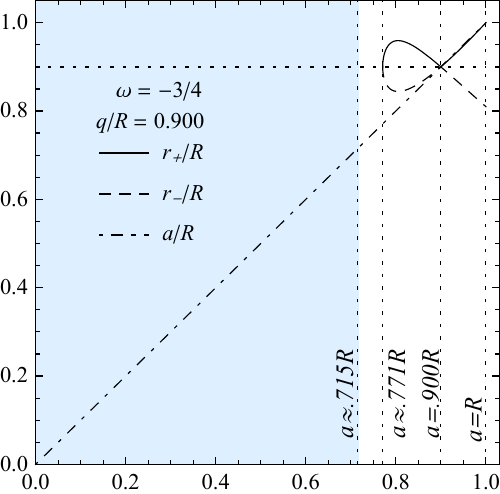}}\\
\subfloat[$\omega=-3/4$; $q/R=1$]{\label{f:w=-34q=1000}
\protect\protect\includegraphics[width=0.38\textwidth]{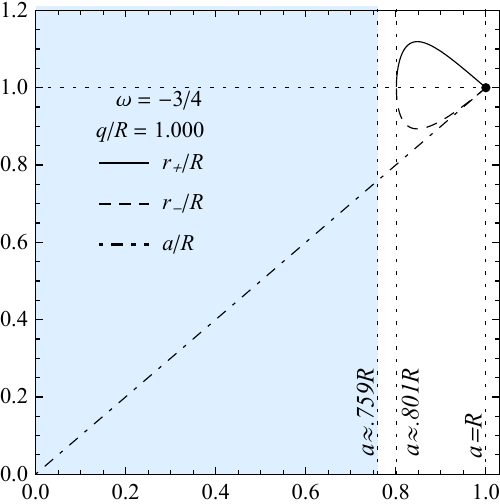}}\quad
\subfloat[$\omega=-3/4$; $q/R=1.400$]{\label{f:w=-34q=1400}
\protect\protect\includegraphics[width=0.38\textwidth]{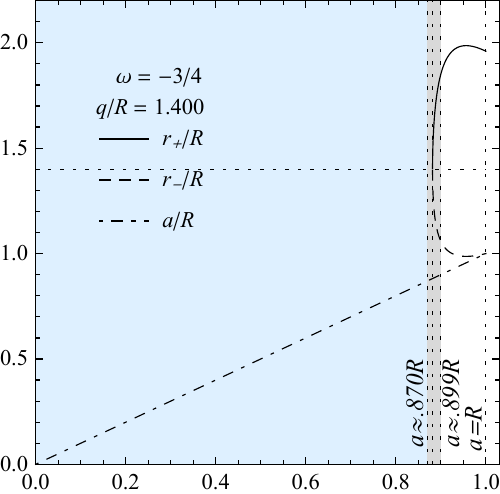}}
\protect\protect\caption{\small Plots of the radii $r_{-}/R$, $r_{+}/R$, and $a/R$ as a
function of $a/R$ for $\omega=-3/4$ and four different values of $q/R$.}
\label{f:w=-34q}
\end{figure*}

\subsubsection{The case  $\omega=-3/4$: Dark energy}

In configurations like this, with negative $\omega\geq-1$, the surface energy-momentum
tensor may be interpreted as a dark energy fluid or a tension shell.
The different kind of compact objects found in this case are shown in Fig.~\ref{f:w=-34q}.

The top left panel of Fig.~\ref{f:w=-34q}, Fig.~\ref{f:w=-34q=0500}, is for $q/R=0.50$ besides $\omega=-3/4$.
In this case the graph shows there different regions.
The interval $0<a/R\lesssim 0.451$ contains no solutions i.e. the mass of the shell assumes complex values.
The solutions in the interval $0.451\lesssim a/R\lesssim0.573$ are
overcharged stars with a thin shell of positive intrinsic mass.
The interval $0.573\lesssim a/R<1$ contains regular undercharged stars. 
The solution at $a/R=1$ is a singular quasiblack hole. The matching radius $a$ approaches $r_+$ from above, the undercharged star becomes more and more compressed and, as seen from Eq.~\eqref{eq:pds}, the shell pressure diverges in that limit.

The top right panel of Fig.~\ref{f:w=-34q}, Fig.~\ref{f:w=-34q=0900}, is for $q/R=0.90$ besides $\omega=-3/4$.
In this case the graph shows four different regions.
The interval $0<a/R\lesssim 0.715$ contains no solutions, i.e., 
the mass of the shell assumes complex values.
The interval $0.715\lesssim a/R\lesssim 0.771$ contains overcharged stars with a thin shell of positive intrinsic mass.
The solution at $a/R\simeq 0.771$ is an extreme regular black hole.
The interval $0.771\lesssim a/R<0.900$ contains regular black holes
with a thin shell of positive mass.
The solution for $a/R=r_+/R=r_-/R= 0.900$ is extreme in the sense that $m_-/R=q/R$, 
and the matching is done on a horizon (a lightlike surface) for the exterior 
spacetime. As in the cases of Figs.~\ref{f:w=34q=0950} and \ref{f:w=0q=0940}, the energy density, the mass and the intrinsic pressure of the shell are all finite quantities in that limit. It may be interpreted as an extreme quasiblack hole.
The interval $0.90 < a/R<1$ contains very compact regular undercharged stars with the matching
surface radius very close to the gravitational radius $a\gtrsim r_+$.
The solution at $a/R=1$ is a singular quasiblack hole. The matching radius $a$ approaches $r_+$ from above and the shell pressure diverges in that limit.

The bottom left panel of Fig.~\ref{f:w=-34q}, Fig.~\ref{f:w=-34q=1000},
is for $q/R=1$ besides $\omega=-3/4$.
In this case the graph shows three different regions.
The interval $0< a/R\lesssim0.759$ contains no solutions, i.e., 
the mass of the shell assumes complex values.
The solutions at $0.759\lesssim a/R\lesssim0.801$ contains overcharged stars with a thin shell of positive intrinsic mass.
The solution at $a/R\simeq 0.801$ is an extreme regular black hole.
The interval $0.801\lesssim a/R<1$ contains regular black holes.
The solution at $a/R=1$ it happens that $a=r_+=r_-=m_-=q$ and the matching surface 
is lightlike. The intrinsic energy density and the mass of the shell
are finite (vanish),
but the intrinsic pressure diverges and then the solution may be interpreted as a singular quasiblack hole.

The bottom right panel of Fig.~\ref{f:w=-34q}, Fig.~\ref{f:w=-34q=1400},
is for $q/R=1.40$ besides $\omega=-3/4$.
In this case the graph shows four different regions.
The interval $0<a/R\lesssim0.870$ contains no solutions, i.e., the mass of
the shell assumes complex values.
The interval $0.870\lesssim a/R\lesssim 0.881$ contains overcharged stars with 
a thin shell of negative intrinsic mass.
The solution at $a/R\simeq 0.881$ is an extreme regular black hole.
The interval  $0.881\lesssim a/R\lesssim 0.899$ contains  regular black holes.
The solution at $a/R\simeq 0.899$ is a regular black hole with no thin shell, 
i.e., the mass of the shell, as well as the intrinsic energy density and pressure are zero.
The interval $0.899\lesssim a/R<1$ contains regular black holes.
The solution at $a/R=1$ is a singular black hole since the  matching happens at $a=r_-$, inside the horizon, and the shell pressure diverges there.

\subsubsection{The case  $\omega=-2$: Phantom matter}

\begin{figure*}[htb]
\centering
\subfloat[$\omega=-2$; $q/R=0.600$]{\label{f:w=-2q=0600}
\protect\protect\includegraphics[width=0.38\textwidth]{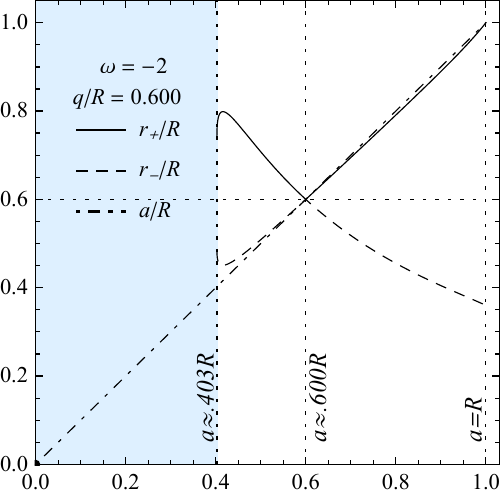}}\quad
\subfloat[$\omega=-2$; $q/R=1.000$]{\label{f:w=-2q=1000}
\protect\protect\includegraphics[width=0.38\textwidth]{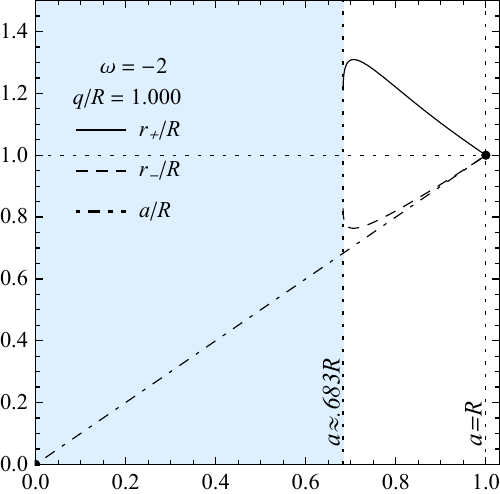}}\\
\subfloat[$\omega=-2$; $q/R=1.300$]{\label{f:w=-2q=1300}
\protect\protect\includegraphics[width=0.38\textwidth]{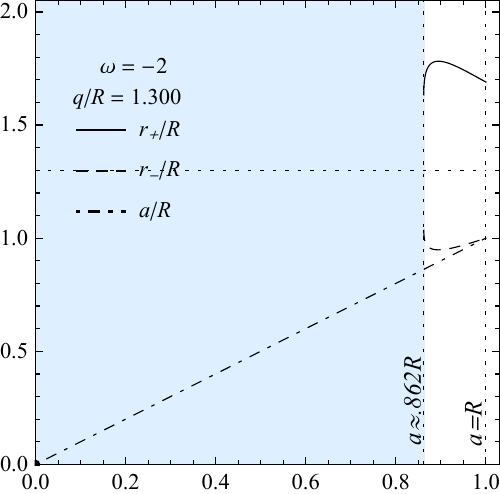}}\quad
\subfloat[$\omega=-2$; $q/R=1.650$]{\label{f:w=-2q=1650}
\protect\protect\includegraphics[width=0.38\textwidth]{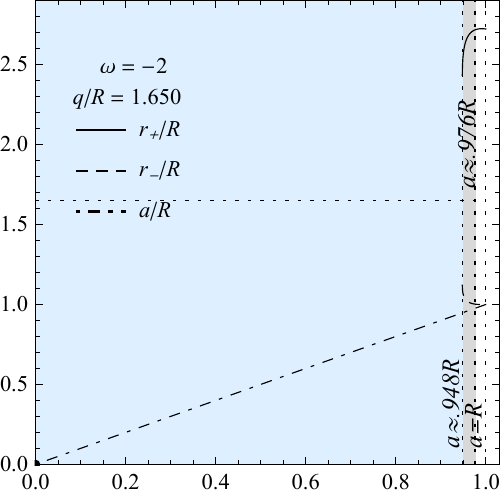}}
\protect\protect\caption{\small Plots of the radii $r_{-}/R$, $r_{+}/R$, and $a/R$ as function 
of $a/R$ for $\omega=-2$ and four different values of $q/R$.}
\label{f:w=-2q}
\end{figure*}

In this configurations, with negative $\omega<-1$, the thin shell carries a fluid 
that may be interpreted as phantom matter. 
The different kind of compact objects found in this case are shown in Fig.~\ref{f:w=-2q}. The properties of the objects spanned in this case may be inferred by comparing to the previous cases of Figs.~\ref{f:w=34q}, \ref{f:w=0q}, and \ref{f:w=-34q}.

The top left panel of Fig.~\ref{f:w=-2q}, Fig.~\ref{f:w=-2q=0600}, 
is for $q/R=0.600$ besides $\omega=-2$.
In this case the graph shows three different regions.
The interval $0< a/R\lesssim 0.403$ contains no physical solutions, i.e., 
the mass of the shell assumes complex values.
The solution at $a/R\simeq 0.403$ is a an extreme regular black hole.
The solutions belonging to the interval $0.403\lesssim a/R< 0.600$ are regular black holes with 
a thin shell of positive intrinsic mass.
The solution at $a/R=r_+/R=r_-/R = 0.600$ bears a shell boundary with finite energy density, mass, and intrinsic pressure. It is an extreme quasiblack hole.
The interval $0.600 < a/R<1$ contains regular undercharged stars with the 
matching surface radius very close to the gravitational radius $a\gtrsim r_+$.
The solution at $a/R=1=r_+/R$ is a singular quasiblack with a shell of infinitely large stress.

The top right panel of Fig.~\ref{f:w=-2q}, Fig.~\ref{f:w=-2q=1000}, is for $q/R=1$ besides $\omega=-2$.
In this case the graph shows two regions.
The interval $0< a/R\lesssim 0.683$ contains no solutions, i.e., the mass of the shell assumes complex values.
The solution at $a/R\simeq 0.683$ is a regular extreme black hole.
The solutions in the interval $0.683\lesssim a/R<1$ are regular black holes with a thin shell of positive intrinsic mass.
The solution at $a/R=1$ also satisfies the relations  $a=r_+=r_-=m_-=q$ and the matching surface is lightlike. 
The solutions is a singular quasiblack hole since the intrinsic energy density and the mass of the shell are finite (vanish), but the intrinsic pressure diverges.

The bottom left panel of Fig.~\ref{f:w=-2q}, Fig.~\ref{f:w=-2q=1300}, 
is for $q/R=1.30$ besides $\omega=-2$.
In this case the graph shows two different regions.
The interval $0< a/R\lesssim 0.862$ contains no solutions, i.e., the mass of the shell assumes complex values.
The solution at $a/R\simeq 0.682$ is a regular extreme black hole.
The solutions in the interval $0.6862\lesssim a/R<1$ are regular black holes  with a thin shell of positive intrinsic mass.
The solution at $a/R=1$ is a singular black hole since the matching is at $a=r_+$ and pressure of the shell is infinitely large. 

The bottom right panel of Fig.~\ref{f:w=-2q}, Fig.~\ref{f:w=-2q=1650} is for $q/R=1.65$ besides $\omega=-2$.
In this case the graph shows three regions.
The interval $0<a/R\lesssim 0.948$ contains no solutions i.e. the mass of the shell assumes complex values.
The solution at $a/R\simeq 0.948$ is a regular extreme black hole with a shell of negative mass.
The solutions in the interval $0.948\lesssim a/R\lesssim 0.976$ are regular black holes with 
a thin shell of negative intrinsic mass.
The solution at $a/R\simeq 0.976$ is a regular black hole with no thin shell, i.e.,
the mass of the shell, as well as the intrinsic energy density and pressure are zero.
The interval  $0.976\lesssim a/R<1$ contains regular black holes.
The solution at $a/R=1$ is a singular black hole since the pressure of the shell, located at the inner horizon $a=r_-$,
diverges.

\subsection{Analysis and classification of solutions in the parameter
space $q/R\times a/R$}

\label{sec:spaceM-}

\subsubsection{General remarks}
For a given $\omega$, each point in the $(q/R,\,a/R)$-plane represents a
specific solution. In Fig. \ref{f:Mq} we plot level curves for $M_{-}$ 
as well as other relevant curves  which 
allow us to single out the point which corresponds to the particular configuration of interest.
Such curves are boundaries of different regions in the two-dimensional parameter space.   
Below we analyze separately the kind of solutions each region and boundary lines represent.

\subsubsection{The line $q/R=0$}

This line is the vertical axis $q/R=0$ in the Fig. \ref{f:Mq}. It represents the limit 
of zero charge and finite radius $a$. The solutions on this line are
composed by the junction of the Schwarzschild exterior solution to a de Sitter-like inner solution.
The radius of the thin shell $a$ is outside the gravitational radius, i.e.,
$a> r_{s}=2m$, hence there are no black hole solutions on this line. Taking this limit, 
the fluid quantities given in Eqs. \eqref{eq:denEnerIn}-\eqref{eq:pretanin} take the form
$\rho_{m}(r)=-p_{r}(r)=-p_{t}(r)=3/R^{2}$. Therefore, the function $\mathcal{M}(r)$
of Eq. \eqref{eq:Massadentro} is 
$\mathcal{M}(r)=r^{3}/\left(2R^{2}\right)$.
In the limit $a \rightarrow 0 $, both $m_-$ and $M_{-}$ vanish and the 
solution at point $M_s$ of Fig.~\ref{f:Mq}  describes a Minkowski spacetime.
On the other hand, in the limit $a/R \rightarrow 1$  the 
intrinsic pressure of the shell diverges and the solution is singular.

\subsubsection{The line $m_-/R=q/R$}
\begin{figure*}[htb]
\centering 
\subfloat[$\omega=3/4$]{\label{f:w=34Mq}\protect\protect\includegraphics[
width=0.40\textwidth]{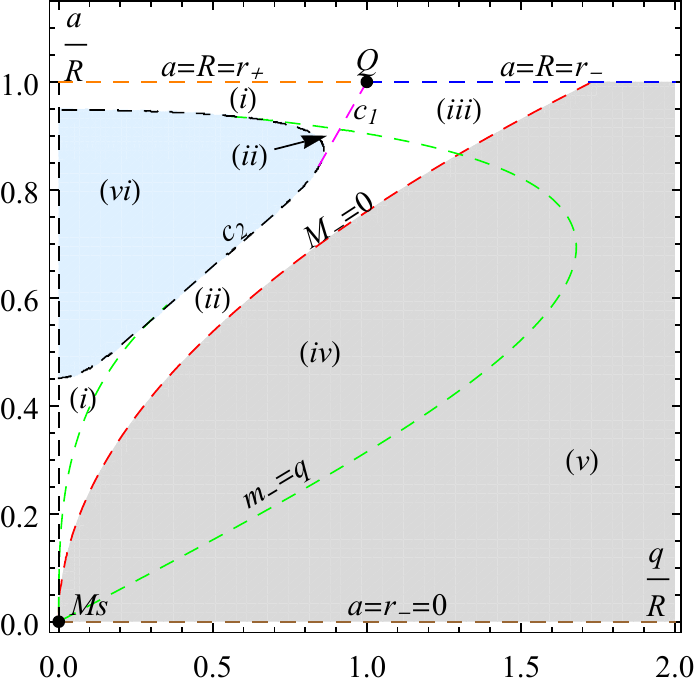}}\,
\subfloat[$\omega=0$]{\label{f:w=0Mq}\protect\protect\includegraphics[
width=0.40\textwidth]{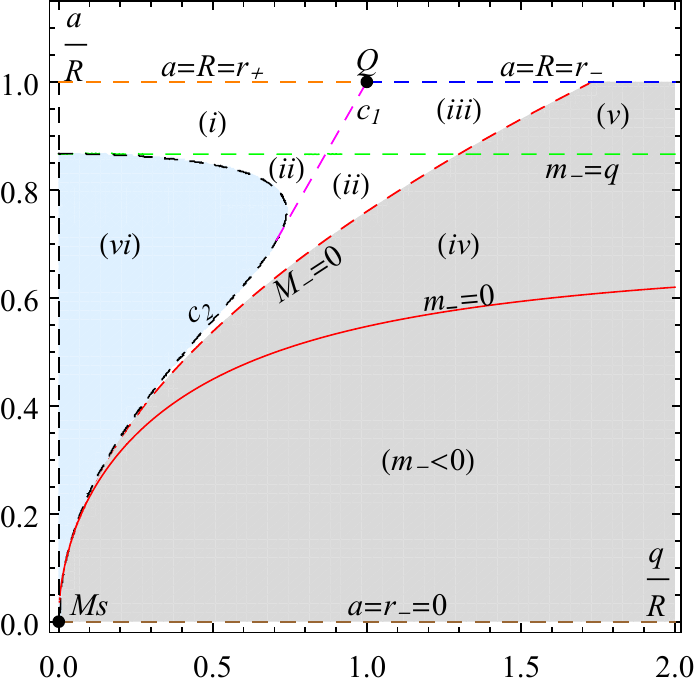}}\\
\subfloat[$\omega=-3/4$]{\label{f:w=-34Mq}\protect\protect\includegraphics[
width=0.40\textwidth]{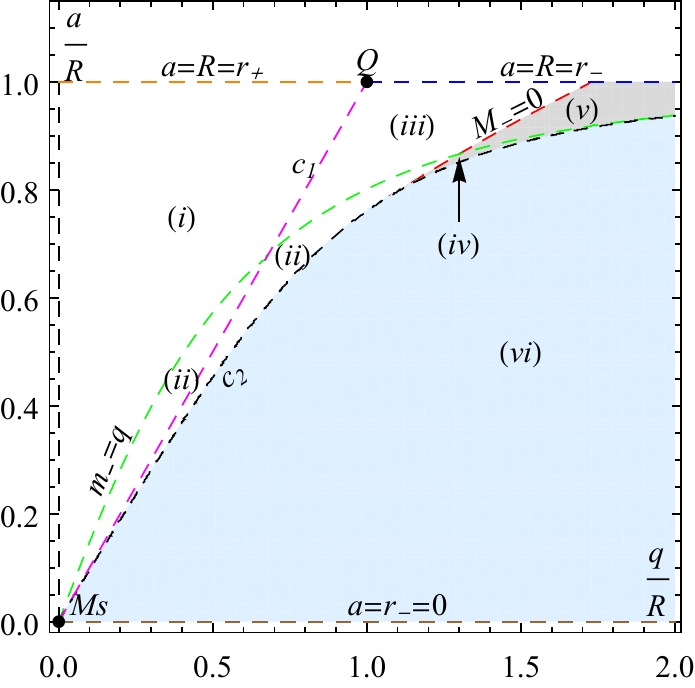}}\,
\subfloat[$\omega=-2$]{\label{f:w=-2Mq}\protect\protect\includegraphics[
width=0.40\textwidth]{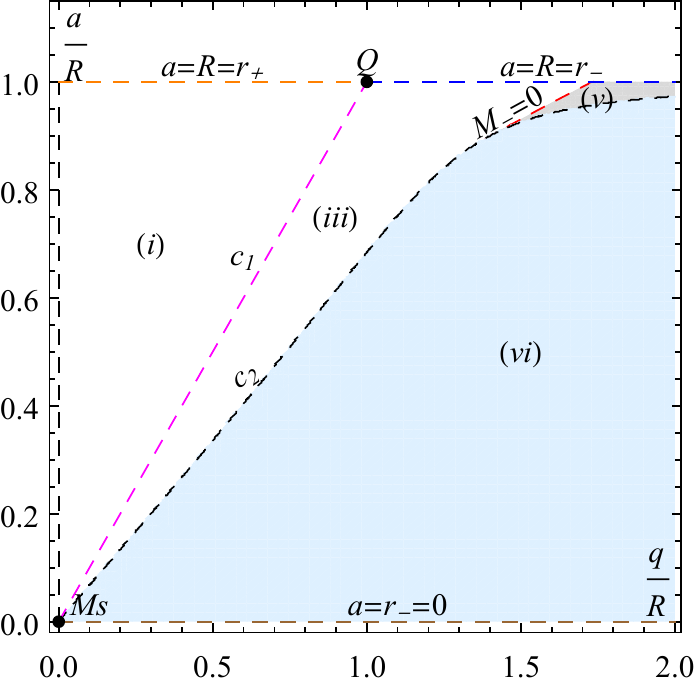}}
\protect\protect\caption{\small\small 
Level curves of $M_{-}/R$ and $m_-/R$ in the parameter space $(q/R,\,a/R)$ for  
four different values of $\omega$, as indicated. 
 Other important curves and the
several different regions are also shown.} 
\label{f:Mq} 
\end{figure*}

 This line contains the extremely charged solutions of the exterior
 Reissner-Nordstr\"om metric, it is obtained by taking $m_-/R=q/R$ into
 Eq. \eqref{eq:totalmass}, and the resulting solutions are represented by dashed green lines in Fig.~\ref{f:Mq}.
 For non-negative $\omega$, equation $m_-(a/R,q/R)/R=  q/R$ gives a closed curve, a branch of which coincides with the curve $c_2$. The internal area of such closed curve represents
 all configurations with  total gravitational mass $m_-$ smaller that total electric charge
 $q$ (overcharged solutions), while the external area characterizes solutions with $m_->q$ 
 (undercharged solutions). The lines for $m_-/R=q/R$ are not closed curves 
 for $\omega \leq 0$, as seen in Figs.~\ref{f:w=0Mq} and \ref{f:w=-34Mq}
 for $\omega=0$ and $\omega =-3/4$, respectively. In these 
 cases the undercharged solutions are in the region above the respective lines,
 while the overcharged solutions and/or the imaginary solutions 
 lie below such lines. As $\omega$ decreases to higher negative values, the curve for $m_-=q$ tends to coincide with $c_2$, and for $\omega \lesssim -1.725$ it is not present.

\subsubsection{The line $M_-/R=0$}

The curve corresponding to $M_-/R=0$ as a function of $a/R$ and $q/R$ is plotted  in Fig.~\ref{f:Mq} by a dashed red line, also indicated by the appropriate label $M_-=0$. 
This line represents solutions without a thin shell at the boundary, because it also
gives $\sigma=0$ and $\mathcal{P}=0$ [cf. Eqs.~\eqref{eq:eds} and  \eqref{eq:pds}]. 
In such a situation, the junction between the interior de Sitter and the exterior Reissner-Nordstr\"om spacetimes is made smoothly, 
see also Refs.~\citen{Lemos:2011vz,Japoes2012}.
The area above this line represents
configurations with $M_-/R>0$, while the area below it represents
configurations with $M_{-}/R<0$. Notice that, for $\omega< -1/4$,
 the mass $M_-/R$ also assumes imaginary values below the 
lines $M_-/R=0$, see regions ($vi$) in Figs.~\ref{f:w=-34Mq} and
\ref{f:w=-2Mq}.
The condition $M_-/R=0$ imposed into Eq.~\eqref{eq:massadashell} furnishes 
two real solutions, $q=\pm {\sqrt{3}a^{2}}/{R}$.
Substituting these relations into Eq. \eqref{eq:totalmass} we find
$m_-={2a^{3}}/{R^{2}}$ together with the additional condition for $\omega> {-a^2}/{2\left(1-a^2\right)}$.
Note that in the limit $a\rightarrow R$ one has
$m_-/R\rightarrow 2$ and $q/R\rightarrow \sqrt{3}$.
In such a limit, the pressure 
also vanish and the solution is similar to the isotropic case studied 
in Ref.~\refcite{Lemos:2011vz}.

\subsubsection{The line $a/R=r_-/R=0$}

One of the solution of the equation $r_{-}/R=a/R$, with $r_-/R$ from relation~\eqref{eq:rpm},  is $a/R=r_-/R=0$, which is satisfied for all $q/R\geq 0$. 
This is the horizontal axis with $q/R\geq0$ and it is a real solution just for
$\omega> -1/4$.  It is plotted in the panels of Fig.~\ref{f:Mq} by dashed brown lines, also indicated by the appropriate label $a=r_-=0$.   More precisely, we find $
\underset{a\rightarrow 0}{\mathrm{lim}}M_{-}=-{|q|}/\sqrt{1+4\omega}$.
Note that for $ \omega>-1/4$ the mass of the shell is negative, while
for  $\omega<-1/4$ it is imaginary. On the other hand, the leading term for the total mass in the limit $a\to 0$ is $m_- = 2 q^2 \omega/(1 + 4 w) a$. Therefore, outside the interval $-1/4<\omega<0$ one gets 
$m_-/R\rightarrow\infty$ and $r_+/R\rightarrow\infty$, besides $r_-/R\to 0$, so that the solutions belonging to this line correspond to Kasner spacetimes (see Appendix A of Ref.~\refcite{Lemos:2017vz} for details).
Within the interval $-1/4<\omega<0$, one has $m_-/R\rightarrow-\infty$ and the solutions are naked singularities. The case $\omega=0$ gives $m_- = M_- =-|q|$ and then the solution is a naked singularity.

\subsubsection{The line $a/R=r_-/R=1$}

Another solution of the equation $r_-/R=a/R$ is $a/R=r_{-}/R=1$, which is satisfied 
for all $q/R> 1$. It is plotted in the panels of Fig.~\ref{f:Mq} by dashed blue lines, also indicated by the appropriate label $a=R=r_-$.
In this solution the junction is made on the inner horizon of the  
Reissner-Nordstr\"om metric and on the horizon of the de Sitter metric. 
It is a limiting case where the intrinsic energy density and the mass of
the shell vanish. However, on this line 
the intrinsic pressure of the shell diverges resulting in a singular solutions. 
The solution is a regular black hole just at $q/R=\sqrt{3}$ for which $m_-/R=2$ 
and the pressure also vanishes. This particular solution
is similar to the one studied in Ref.~\refcite{Lemos:2011vz}.

\subsubsection{The line $a/R=r_+/R=1$}

In this case one has $a=m_-+\sqrt{m_-^{2}-q^{2}}$ whose solution is  $a/R=r_{+}/R=1$, and that is satisfied for all $0<q/R<1$. The corresponding solution is represented in the panels of Fig.~\ref{f:Mq} by dashed yellow lines, also indicated by the appropriate label $a=R=r_+$.
The junction is made on the exterior
horizon of the Reissner-Nordstr\"om metric and on the horizon of the de Sitter
metric. This line contains singular solutions since the intrinsic pressure of the shell diverges and may be interpreted as nonextreme quasiblack holes\cite{Lemos:2007yh}, see also Ref.~\refcite{Barrabes1991}.

\subsubsection{The line $m_-=0$}
This line separates the parameter space into a region containing physically interesting objects with positive and other with negative gravitational mass. The line is present just in a finite interval of $\omega$, namely, for $\omega\in (-1/2,\, 1/2)$. For $\omega$ in the interval $(0,\,1/2)$ the solution to the equation $m_-(a/R, q/R)=0$ is a closed curve, with the positive mass solutions being in the exterior region.
In the case $\omega =0$, depicted in Fig.~\ref{f:w=0Mq}, the curve closes in the region $q/R\to \infty$. For $\omega$ in the interval $(-1/2,\,0)$ the curve presents two disconnected branches, one of them being similar to the case of $\omega=0$ shown in Fig.~\ref{f:w=0Mq}, except that it does not extend to $a/R\to 0$. In this region of small $q/R$, the solution is given by the second branch. The positive mass solutions are located in the region above such curves.   

\subsubsection{The line $c_1$}

This line is drawn for the condition $r_-/R=a/R$ and the solution is a segment of the line $a/R=q/R$. It is represented in the graphs of Fig.~\ref{f:Mq} by dashed magenta lines, also indicated by the appropriate label $c_1$. 
Notice that, for $\omega> -1/2$, $c_1$ has one of its endpoints on the dashed black $c_2$ line (drawn for $M_-=M_+$), as seen in Figs.~\ref{f:w=34Mq} and \ref{f:w=0Mq}. The other endpoint is at $a/R=1$, represented by the point $Q$ in all panels of Fig.~\ref{f:Mq}. As seen from the figures, starting at point $Q$, a first segment of this line is the boundary between region $(i)$ of regular black holes and region $(iii)$ of regular undercharged stars, while a second segment penetrates region $(ii)$, when present, splitting it into two parts, and reaches curve $c_2$.
For $\omega \leq -1/2$ the line is complete in the diagram, connecting the point $M_s$ to $Q$, cf. Figs.~\ref{f:w=-34Mq} and \ref{f:w=-2Mq}.  
All the quantities $M_-$, $m_-$, $\sigma$, and $\mathcal{P}$ are well defined on this line, except at the endpoint at $a/R=1$ where the pressure of the shell is arbitrarily large. 
Since $m_-=q$ for all the configurations on the line $c_1$, the matching surface is at the extreme horizon ($a=r_-=r_+$) of the exterior Reissner-Nordstr\"om metric. A possible interpretation for these configurations is a kind of extreme quasiblack hole \cite{Lemos:2007yh}. The particular configurations represented by the line $c_1$ were noticed in the descriptions of particular cases presented in Sect.~\ref{sec:radii-1}. The configurations at the endpoints $M_s$ and $Q$ are discussed below.  

\subsubsection{The line $c_2$}

This line represents the boundary of real solutions for the thin shell mass $M_-$. 
It was drawn by putting the imaginary part of $M_-$ to zero.
All the physical quantities are well defined on such a line, and
then it contains interesting physical configurations whose properties are related to the solutions belonging to the neighbouring regions.

\subsubsection{The point $M_s$}

The point $M_s$ is the limit $(a,\,q) \rightarrow (0,\,0)$ and it represents different solutions, 
since functions $\sigma$, $\cal P$, $M_-$, etc., may assume diferent values in such a limit, depending on the ratio $q^2/a^2$. If the limit is taken along a path for 
which $q^2/a^2$ tends to a finite value, all the quantities $M_-/R$, 
$\mathcal P$, and $m_-/R$ vanish and the solution represents the Minkowski spacetime. On the other hand, if the path is such that 
$q^2/a^2$ diverges in the limit $(a/R,\,q/R) \rightarrow (0,\,0)$, the
solution is a Kasner spacetime. See Ref.~\refcite{Lemos:2017vz} for more details on similar situations.

\subsubsection{The point $Q$}

 The point $Q$ is a degenerate point in the diagrams, being the convergence point
 of three special lines, namely, $c_1$, $a/R=r_-/R$, and $a/R=r_+/R$.
 Here, all the quantities $a/R$, $q/R$, $m_-/R$, $r_+/R$, and $r_-/R$ have the same value,
 they all equal unity. The intrinsic energy density and the mass of the
 shell have zero limiting values, while the intrinsic pressure diverges for 
 all $\omega$. It represents a solution with a regular interior but with 
 a singular coat of matter at the boundary. It may be interpreted as a singular quasiblack hole\cite{Lemos:2007yh}.

\subsubsection{The region $(i)$}

This is the region of regular charged star configurations with total mass 
larger than the total electric charge, $m_-/R>q/R$, see Fig. \ref{f:Mq}.
It is delimited by the vertical axis $q/R=0$ and by the curves $c_{1}$, 
$c_{2}$, $m_-/R=q/R$, and the horizontal line $a/R=r_+/R=1$. In this region the radius of the shell $a/R$
satisfies the constraint $a/R>r_{+}/R$, so there is no black holes and all solutions are regular stars with a de Sitter core and a surface layer at
the boundary.
These stars have positive total mass $m_-$ and a thin shell of positive mass $M_-$ whose values depend also on $\omega$, see Eq. \eqref{eq:massadashell}, and it is present for all $\omega$.

\subsubsection{The region $(ii)$}

This is the region of overcharged star configurations with total mass 
smaller than the total electric charge,  $m_-/R< q/R$, see Fig. \ref{f:Mq}. 
It is delimited by the curves $c_{2}$, $m_-=q$, and $M_{-}=0$, and it is 
present for all $\omega\gtrsim-1.725$,  see for  instance the cases $\omega =3/4$ ([panel (a)], $\omega= 0$ [panel (b)], and $\omega=-3/4$ [panel(c)] in Fig. \ref{f:Mq}.
It is not present for $\omega\lesssim -1.725$, see panel (d) of Fig. \ref{f:Mq}. 
In this region the radius of the shell $a/R$
satisfies the constraint $a/R>r_{+}/R$, so there is no black holes and  
all solutions are regular stars with a de Sitter core and a thin shell of non-negative mass
$M_-/R\geq 0$, $M_-/R=0$ being part of the region boundary. The class of overcharged stars belonging to this region satisfies the following 
constraint for the total electric charge $q/R$,
$
0<{q}/{R}<3\sqrt{3}/{4}$.
The upper bound value $q/R=3\sqrt{3}/4$ and the condition $M_-=0$ 
furnish the relation $a/R=\sqrt{3}/2$. 
The point ($q/R=3\sqrt{3}/4,\, a/R=\sqrt{3}/2$) represents the intercept between 
the line $m_-/R=q/R$ and the line $M_{-}/R=0$.

\subsubsection{The region $(iii)$ }

This is the region of regular black holes with a de Sitter core
and a thin shell at the boundary of the core. 
For $\omega \gtrsim -1.725$, the region is delimited by the lines $c_1$, $m_-/R=q/R$, $M_{-}/R=0$, 
and the line $a/R=r_{-}/R=1$, see panels (a), (b) and (c) of Fig. \ref{f:Mq}.  For $\omega\lesssim -1.725$, region $(iii)$ is delimited by the lines 
$c_1$, $c_2$, $a=R=r_-$, and the curve $M_-/R=0$, see panel (d) of Fig. \ref{f:Mq}. 
All objects contained in such a region satisfy the constraint $a/R<r_{-}/R$, 
confirming they are all regular black holes.
 Moreover, all solutions in this region have a large charge to mas ratio, i.e.,
 the solutions satisfy the condition $\sqrt{3}/{2}<{q}/{m_-}\leq 1 $ for all $\omega$.

\subsubsection{The region $(iv)$}
This is the region of overcharged stars with $m_-/R<q/R$ with a thin shell of negative mass $M_-/R<0$ for all solutions. The region is delimited 
by the lines $M_-/R=0$ and $m_-/R=q/R$ for $\omega >0$, cf. panel (a) of Fig.~\ref{f:Mq}, by  the lines $M_-/R=0$, $m_-/R=q/R$
and by the horizontal line $a/R=r_-/R=0$ for $\omega=0$,  and it 
is not present for all $\omega \leq -1.5$.
Notice that region $(iv)$ is present in Fig.~\ref{f:w=-34Mq} for $\omega =-3/4$,
but it is absent in Fig.~\ref{f:w=-2Mq} for $\omega =-2$.
In this region the radius of the shell $a/R$
satisfies the constraint $a/R>r_{+}/R$, so there is no black holes and all solutions are overcharged stars with a de Sitter core and a surface layer at
the boundary.
 Moreover,
it can be shown that for $-1/2\leq\omega\leq 1/2$ there are regions
of negative total gravitational mass, $m_-/R<0$. These solutions are then of 
little interest.

\subsubsection{The region $(v)$}

This is another region of regular black hole solutions. 
The total mass is positive and satisfies $m_-> q$, but the thin shell mass $M_-$ is negative. 
The region is delimited by different lines depending on $\omega$: 
by the horizontal lines $a/R=r_{+}/R=1$ and $a/R=r_{-}/R=0$, the 
line $M_{-}/R=0$, 
and the curve $m_-/R=q/R$ for $\omega> 0$, cf. panel (a) of Fig.~\ref{f:Mq}; by the lines  $M_{-}/R=0$, $a/R=r_{+}/R=1$,  
and $m_-/R=q/R$ for $\omega>-1.5$, cf. panels (b) and (c) of Fig. \ref{f:Mq}; and by the lines $M_{-}/R=0$, $a/R=r_{+}/R=1$,  
and $c_2$ for  $\omega\leq -1.5$, cf. Fig. \ref{f:w=-2Mq}.

\subsubsection{The region $(vi)$}
 This is a region with no real solution for $M_{-}$, i.e., $M_-$ assumes 
 complex values meaning that there are no physical configurations. 
 The region is delimited by the vertical line $q/R=0$ and the curve $c_{2}$ for $\omega> -1/4$, 
 and by the curve $c_2$ and the horizontal line $a/R=r_{-}/R=0$ for $\omega <-1/4$, see Fig. \ref{f:Mq}.

\section{Second class of solutions: for $M=M_+$ and $m=m_+$}
\label{secondclass}

\subsection{General remarks}
\label{sec:generalM+}

The second solution for the shell mass, $M=M_+$, is obtained by taking the
plus sign in Eq.~\eqref{eq:massadashell}, which corresponds also to the 
plus sign solution in Eq.~\eqref{eq:totalmass}. 
the total mass of the system $m_+$ assumes also negative values in a range of parameters.

The singular points in the parameter space are the same as in the case for
$M_-$,  i.e., $a/R=1$, and $\omega = -1/4$, see Secs.~\ref{sec:prelim} and \ref{sec:generalM-}.
The case for  $\omega = -1/4$ was considered in Sec.~\ref{sec:prelim}.
When the radius of the junction surface $a$ coincides with $R$, the plus 
sign solution in Eq. 
\eqref{eq:massadashell}  is singular. In fact, taking the limit
$a\rightarrow R$ we find
$ \underset{a \rightarrow R}{\mathrm{lim}}M_+/R= \pm \infty,$
which means that if the matching surface is placed at $a/R=1$  it results in singular solutions.

Below we analyze some interesting properties of this second solution for $M$.
For the sake of comparison we take the same values of the equation of state 
parameter $\omega$ as in the case for $M_-$.

\begin{figure*}[htb]
\centering
\subfloat[$\omega=3/4$; $q/R=0.400$]{\label{f:w=34M2q=0400}
\protect\protect\includegraphics[width=0.38\textwidth]{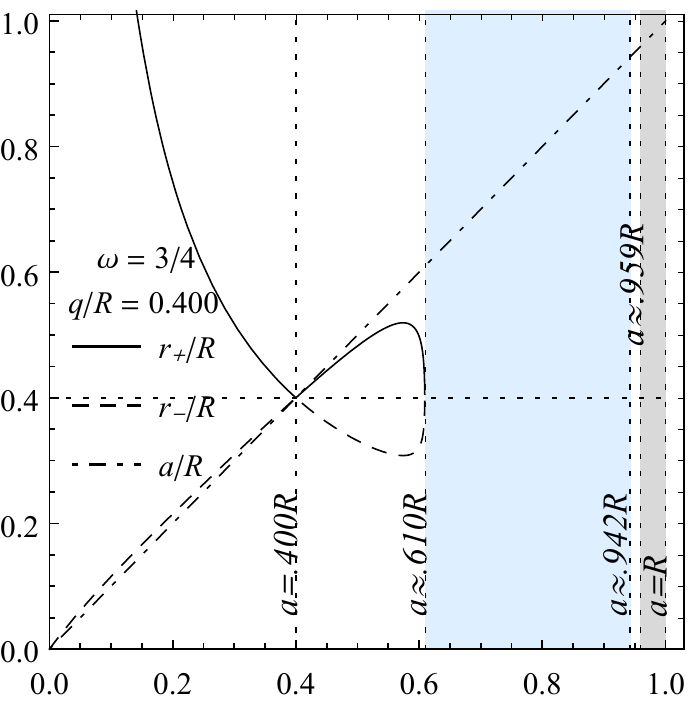}}\,
\subfloat[$\omega=3/4$; $q/R=0.650$]{\label{f:w=34M2q=0650}
\protect\protect\includegraphics[width=0.38\textwidth]{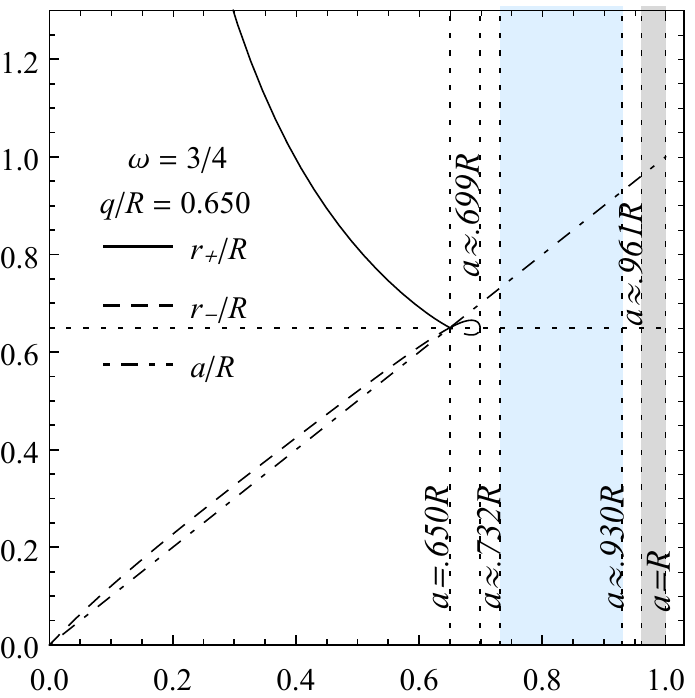}}\\
\subfloat[$\omega=3/4$; $q/R=1.000$]{\label{f:w=34M2q=1000}
\protect\protect\includegraphics[width=0.38\textwidth]{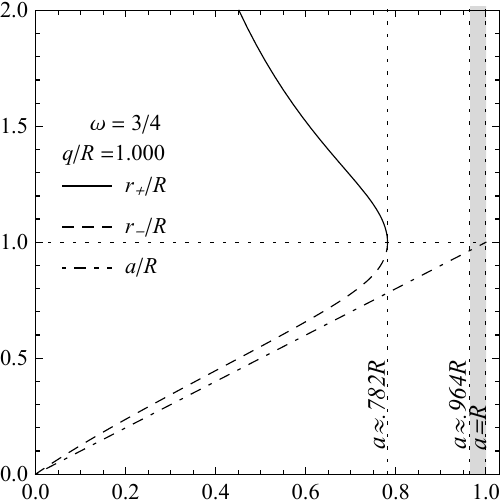}}\,
\subfloat[$\omega=3/4$; $q/R=1.400$]{\label{f:w=34M2q=1400}
\protect\protect\includegraphics[width=0.38\textwidth]{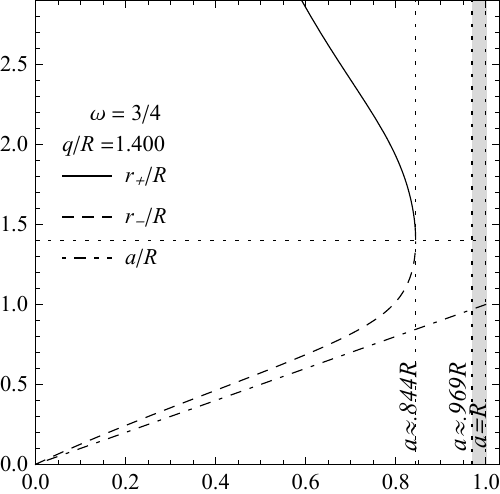}}\,
\protect\protect\caption{\small Plots of the radii $r_{-}/R$, $r_{+}/R$, and $a/R$
as a function of $a/R$ for $\omega=3/4$ and four different values of $q/R$, as indicated.}
\label{f:w=34M2q2}
\end{figure*}

\subsection{Analysis of the radius $r_{+}$ and $r_{-}$}
\label{sec:radii-2}

\subsubsection{General remarks}

In this subsection we make for the second solution $M_+$ a similar 
analyses as done in Sect. \ref{sec:radii-1} for the first solution $M_-$.
The masses $M_+=M_+(a/R,\,q/R,\,\omega)$ and $m_+=m_+(a/R,\,q/R,\,\omega)$ are 
defined respectively by Eqs. \eqref{eq:massadashell} and \eqref{eq:totalmass} .

 In the following we plot the radii $r_{+}/R$, $r_{-}/R$, and $a/R$ 
as a function of $a/R$, for the same values of $\omega$ and the total charge $q/R$ 
considered in Sect.~\ref{sec:radii-1}. 
Later on we analyze the solutions in the parameter space $(q/R,\,a/R)$.
The chosen values of $\omega$ are representative of the wide classes of solutions
that can be found in the present model.

\subsubsection{The case $ \omega = 3/4$}

For this particular solution of $M_+$, the radii $r_+/R$ and $r_-/R$ 
as a function of $a/R$, and for four different values of the
electric charge, are plot in Fig. \ref{f:w=34M2q2}. 

The top left panel of  Fig. \ref{f:w=34M2q2}, Fig.~\ref{f:w=34M2q=0400}, is for $q/R=0.400$ 
besides $\omega=3/4$. It presents five different regions.
The interval $0<a/R < 0.400$ contains regular black holes with a thin shell at the boundary which is located inside the Cauchy horizon.
The solution at $a/R=0.400$ is an extreme quasiblack black hole.
The interval $0.400 < a/R\lesssim 0.610$ contains regular charged stars satisfying $m_+>q$.
The solution at $a/R\simeq 0.610$ is an extremely charged star.
The interval $0.610\lesssim a/R\lesssim 0.942$ contains no physical solutions, the total mass assumes complex values. 
The interval $0.942\lesssim a/R\lesssim 0.959$ contains overcharged stars. 
The solution at $a/R\simeq 0.959$ has zero total mass.
The interval $0.959\lesssim a/R\leq 1$ contains overcharged stars with positive thin shell 
mass at the boundary, but the total mass $m_+$ is negative and the solutions are of little interest.

The top right panel of  Fig. \ref{f:w=34M2q2}, Fig.~\ref{f:w=34M2q=0650}, is for $q/R=0.650$ 
besides $\omega=3/4$. It presents six different regions.
The interval $0<a/R < 0.650$ contains regular black holes with a thin shell at the boundary which is located inside the Cauchy horizon.
The solution at $a/R=0.650$ is an extreme quasiblack black hole.
The interval $0.650 < a/R\lesssim 0.690$ contains regular charged stars satisfying $m_+>q$.
The solution at $a/R\simeq 0.690$ is an extremely charged star.
The intervals $0.690\lesssim a/R\lesssim 0.732$ and $0.933\lesssim a/R\lesssim 0.961$ contain overcharged stars. 
The interval $0.732\lesssim a/R\lesssim 0.933$ contains no physical solutions, the total mass assumes complex values. 
The solution at $a/R\simeq 0.961$ has zero total mass.
The interval $0.961\lesssim a/R\leq 1$ contains overcharged stars with positive thin shell 
mass at the boundary, but the total mass $m_+$ is negative and the solutions are of little interest.

The bottom left panel of  Fig. \ref{f:w=34M2q2}, Fig.~\ref{f:w=34M2q=1000}, is for $q/R=1.000$ 
besides $\omega=3/4$. It presents just three different regions.
The  interval $0<a/R\lesssim 0.782$ contains regular blach holes.
The solution at $a/R\simeq 0.782$ is a regular extreme black hole.
The interval $0.782\lesssim a/R\lesssim 0.964$ contains overcharged stars.
The solution at $a/R\simeq 0.964$ has zero total mass.
The interval $0.964\lesssim a/R\leq 1$ contains overcharged stars with positive thin shell 
mass at the boundary, but the total mass $m_+$ is negative and the solutions are of little interest.

The bottom right panel of Fig. \ref{f:w=34M2q2}, Fig.~\ref{f:w=34M2q=1400}, is for $q/R=1.400$ 
besides $\omega=3/4$. It also presents just three different regions. The situation here is similar to
the case of Fig.~\ref{f:w=34M2q=1000} and then we dot not need to run into details.

All solutions for this particular case with $\omega =3/4$ present positive thin shell mass, $M_{+}/R>0$.

\subsubsection{The case $\omega=0$}

In this case, neglecting solutions with thin shells carrying negative mass, we find only overcharged star
solutions for all values of the electric charge $q/R$, and then we do not show figures
for the radii $r_+$ and $r_-$ here.
More details on this case are given in the next subsection.

\subsubsection{The case $\omega =-3/4$}

In this case, neglecting solutions with thin shells carrying negative mass, we find 
only overcharged star solutions. These objects are found for values of the electric 
charge in the interval $0<q/R\lesssim 1.039$, and for 
the radius of the shell in the interval $0<a/R\lesssim 0.774$, and hence it is not 
necessary to show graphs for the radii $r_+$ and $r_-$ here.
More details on this case are given in the next subsection.

\subsubsection{The case $\omega =-2$}

In this case, neglecting solutions with thin shells carrying negative mass, 
we find only overcharged stars solutions  and then we do not show figures 
for the radii $r_+$ and $r_-$ here. Such stars are found in the interval 
$0<q/R\lesssim 1.386$ for the electric charge, 
with the shell radius being in the interval $0<a/R\lesssim 0.894$.
More details on this case are given in the next subsection.

\subsection{Analysis and classification of solutions in the parameter
space $q/R\times a/R$}

\subsubsection{General remarks}

\begin{figure*}[htb]
\centering 
\subfloat[$\omega=3/4$]{\label{f:w=34M2q}\protect\includegraphics[
width=0.40\textwidth]{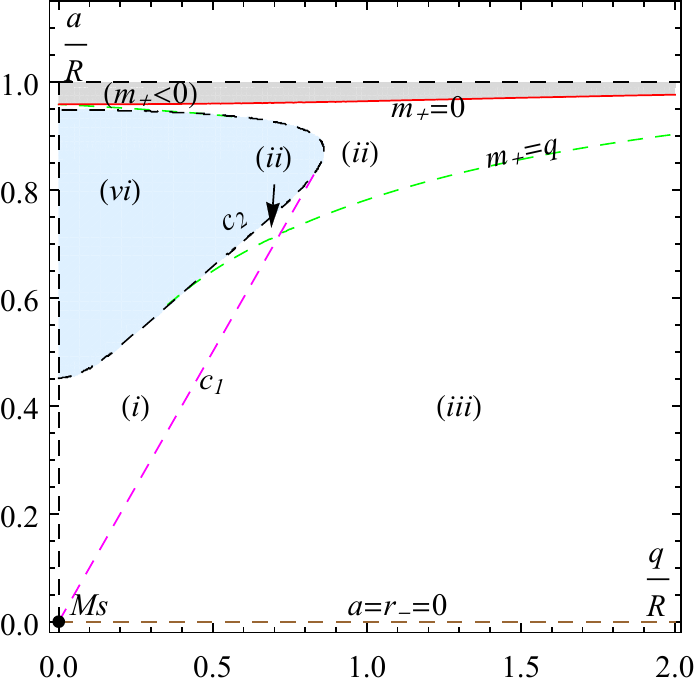}}\,
\subfloat[$\omega=0$]{\label{f:w=0M2q}\protect\includegraphics[
width=0.40\textwidth]{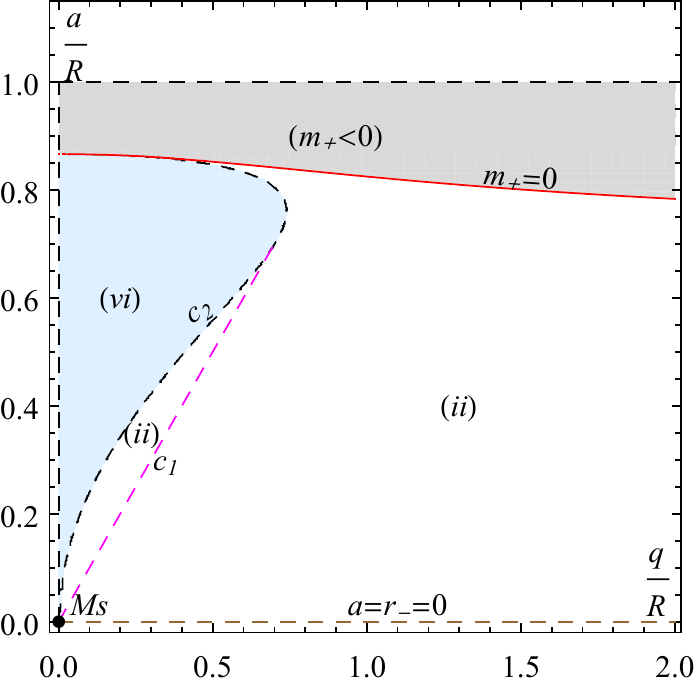}}\\
\subfloat[$\omega=-3/4$]{\label{f:w=-34M2q}\protect\includegraphics[
width=0.40\textwidth]{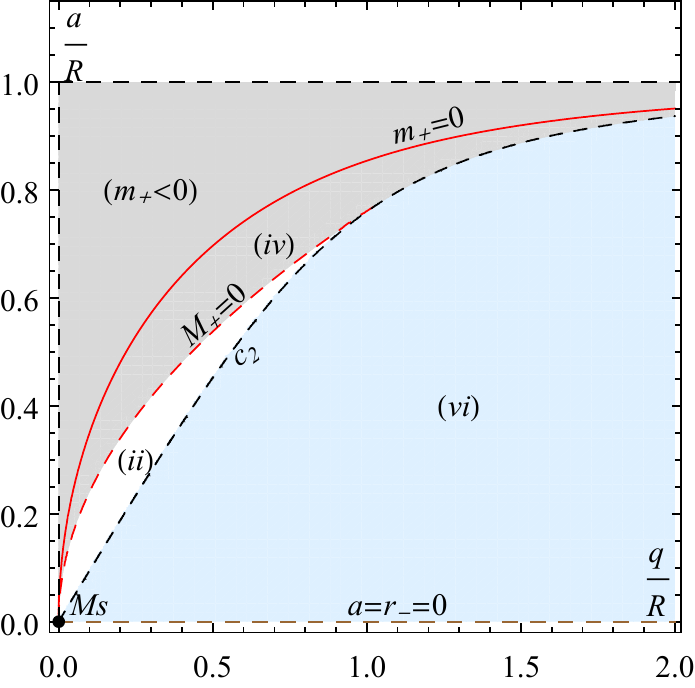}}\,
\subfloat[$\omega=-2$]{\label{f:w=-2M2q}\protect\includegraphics[
width=0.40\textwidth]{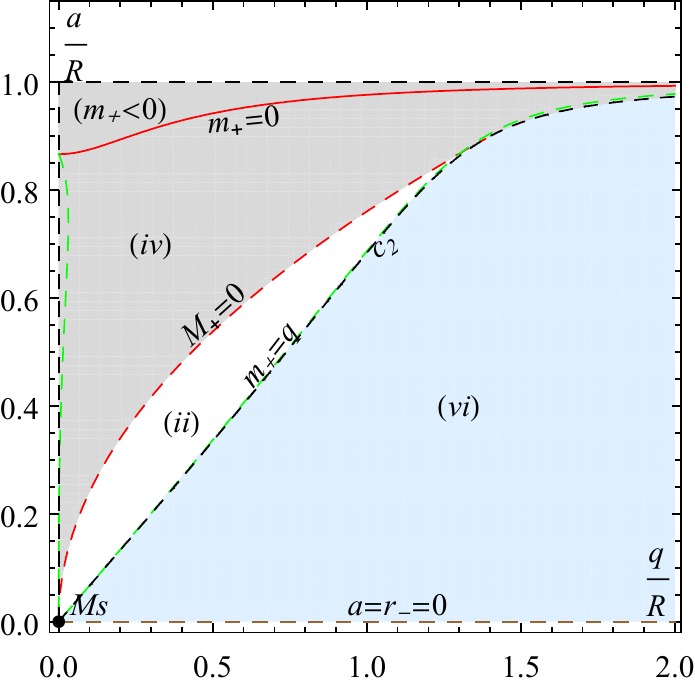}}
\protect\caption{\small\small 
Level curves of $M_+/R$ and $m_+/R$ in the parameter space $(q/R,\,a/R)$ for four different values of $\omega$ as indicated.  
Other important curves and the
several different regions are also shown. }
\label{f:M2q} 
\end{figure*}

In  Fig. \ref{f:M2q} we plot level curves for the second solution of shell mass, $M_{+}$, which 
allows one to single out the point which correspond to physical or unphysical configurations. 
The physical configurations we find are undercharged stars, overcharged stars, and regular black holes, 
all with a de Sitter core and a thin shell of matter at the boundary.
The unphysical solutions are the solutions with negative total mass 
$m_+$ and complex (imaginary) solutions. 
The boundaries of these sectors are lines and points that also represent interesting solutions and we analyse next. The convention and notation for the figures are the same as in the last section.

\subsubsection{The line $q/R=0$}

This is the limit of zero charge and finite radius $a$, i.e., the vertical 
axes of all figures in the parameter space $(q/R,\, a/R)$, see Fig. \ref{f:M2q}. 
The corresponding solutions are composed by the junction of a de Sitter interior region to
the Schwarszchild exterior solution.
The radius of the thin shell $a$ is outside the gravitational radius $r_{s}/R=2m_+/R$ and then all the solutions on this line represent uncharged stars. 
In the limit where the radius of the junction goes to zero, $a/R\to 0$, the masses 
$m_+/R$ and $M_{+}/R$ go to zero too. 
Hence, the resulting solution by approaching the point $M_s$ along this line is the Minkowski metric.
In the limit  $a/R\to 1$ the physical quantities diverge and the solution 
at the point ($q/R=0,\, a/R=1$) is singular.

\subsubsection{The line $m_+/R=q/R$}

 This solution is obtained by putting $m_+/R=q/R$ into Eq. \eqref{eq:totalmass} 
 and using the plus sign solution from  Eq. \eqref{eq:massadashell}. 
This curve have two branches whose form depend strongly on $\omega$, as seen in Fig.~\ref{f:M2q}. One of them, when present, coincides with the curve $c_2$ as seen in Figs.~\ref{f:w=34M2q} and \ref{f:w=-2M2q}. The other branch is not present for $-1< \omega \leq 0 $.
Below this curve the mass $m_+$  is larger than $q$, while above it the mass $m_+$ is smaller than $q$. When the line is not present, one finds just overcharged solutions.

\subsubsection{The line $M_+/R=0$}

 This line represents solutions with no thin shell, in which the junction is made smoothly. 
 The condition $M_+/R=0$ is only satisfied for $\omega<0$, see Fig.~\ref{f:M2q}.
 In such cases,  Eqs. \eqref{eq:massadashell} and \eqref{eq:totalmass} furnish
 the relations $q=\sqrt{3}a^{2}/R$ and $m_+=2a^{3}/R^{2}$.

\subsubsection{The line $a/R=r_-/R=0$}
Similarly to the case of the last section, equation $a/R=r_-/R$ has a solution at the horizontal axes $a/R=0$ that is satisfied for all $q/R>0$. In this case, the thin shell mass is given by
$
\underset{a\rightarrow 0}{\mathrm{lim}}M_{+}={|q|}/\left(\sqrt{1+4\omega}\right)$, which is positive for $\omega>-1/4$, but that is imaginary for $\omega<-1/4$.
Outside the interval $-1/4<\omega<0$, one gets 
$m_+/R\rightarrow\infty$ and $r_+/R\rightarrow\infty$. Hence, the solutions belonging to this line correspond
to Kasner spacetimes, see Appendix A of Ref.~\refcite{Lemos:2017vz} for more details.  
Within the interval $-1/4<\omega\leq 0$, one has $m_+/R\rightarrow -\infty$ and  and the the solution is a naked singularity.

\subsubsection{The line $a/R=1$}
This line corresponds to singular objects with negative total gravitational mass of little interest and then we do not analyse it here. 

\subsubsection{The line $m_+=0$}
This line separates the parameter space into a region containing physically interesting objects with positive gravitational mass, below such a curve, and a region of objects with negative total mass, above such a curve. It is present for all $\omega$, see Fig.~\ref{f:M2q}.

\subsubsection{The line $c_1$}

This line is drawn for the condition $r_-/R=a/R$, whose solution is a segment of the line $a/R=q/R$, starting at point $M_s$ and finishing at the curve $c_2$. The line $c_1$ is present just for $\omega>-1/2$, see Fig.~\ref{f:M2q}. For $\omega> 0$, a segment of $c_1$ is the boundary between the region $(i)$ of regular undercharged stars and the region $(iii)$ of regular black holes, while other segment enters region $(ii)$ and reaches the curve $c_2$. For $-1/2 < \omega\leq 0$, regions $(i)$ and $(iii)$ do not show up
and the line $c_1$ divides the region $(ii)$ of overcharged stars into two parts. Configurations belonging to $c_1$ also satisfy the relation $m_+/R=q/R$, so that the matching is made on the extreme horizon of the exterior Reissner-Nordstr\"om metric, with the corresponding solutions being interpreted as extreme quasiblack holes.

\subsubsection{The line $c_2$}

This line represents the boundary of real solutions for the thin shell mass $M_{+}$. All points on the line $c_{2}$ represent real solutions whose interpretation depends
on the specific values of the parameters, see Fig.~\ref{f:M2q}.

\subsubsection{The point $M_s$}

The point $M_s$ is the limit $(a,\,q) \rightarrow (0,\,0)$ and represents the 
Minkowski spacetime since all the quantities $a/R$, $q/R$, $M_+/R$ and $m_+/R$ 
vanish in such a limit.

\subsubsection{The region  $(i)$ }

This is the region of undercharged star configurations with total mass $m_+/R$ 
larger than the total electric charge $q/R$. 
It is delimited by the curves $c_{1}$, curve $c_{2}$, $m_+/R=q/R$, and vertical axis $q/R=0$, 
and it is present only for $\omega>0$.
In this region the radius $a$ satisfies the constraint $a/R>r_{+}/R$, so the solutions 
all are regular undercharged stars with a de Sitter core and a thin shell of positive mass 
$M_{+}/R >0$.

\subsubsection{The region  $(ii)$ }

This is the region of overcharged star configurations with total mass $m_+/R$ smaller
than the total electric charge $q/R$, see Fig. \ref{f:M2q}. 
For $\omega>0$, it is delimited by the curves $m_+/R=q/R$, $m_+/R=0$ and $c_{2}$. 
For $\omega=0$, it is delimited by the curves $c_{2}$, $m_+/R=0$ and the line $a/R=0$.
Finally, for $\omega< -1/4$, it is delimited by $M_{+}/R=0$ and $c_{2}$. All solutions are regular
 with a de Sitter core and a thin shell of positive mass, $M_{+}/R\geq 0$.

\subsubsection{The region  $(iii)$ }

This is the region of regular black holes with a de Sitter core and positive masses $M_+$ and $m_+$.
It is present for all $\omega> -1/4$. 
It is delimited by the lines $c_{1}$, $m_+/R=q/R$ and $a/R=0$, see Fig. \ref{f:M2q}.

\subsubsection{The region  $(iv)$ }

This is the region of overcharged stars with $m_+/R<q/R$ and a thin shell of negative mass, $M_{+}/R<0$. 
In a range of the radius $a/R$ within such a region also the total mass $m_+$ assumes negative values, 
see Fig.~\ref{f:M2q}.

\subsubsection{The region  $(vi)$ }

This is the region where no real solutions for $M_{+}$ are found, i.e., $M_+$ assumes complex values and
there are no physical configurations. 
For $\omega> -1/4$, the region is delimited by the vertical line $q/R=0$ and the curve $c_{2}$, 
while for $\omega< -1/4$ it is 
delimited by the curve $c_{2}$ and the horizontal lines $a/R=0$ and $a/R=1$,
see Fig. \ref{f:M2q}.

\section{Conclusions}\label{Sec. 4}

We built simple models of regular black holes, quasiblack holes, charged stars 
and other compact objects with charged matter distribution at the core and 
a surface layer joining the core to the electrovacuum exterior.  The models were 
constructed by joining the de Sitter metric (interior) to 
the Reissner-Nordstr\"om metric (exterior) by applying the Darmois-Israel formalism in a 
spherically symmetric spacetime. The distribution of matter in the interior 
is constituted by an electrically charged non-isotropic fluid with radial 
pressure $p_{r}$ satisfying a de Sitter equation of state
$p_r = -\rho_m$, where $\rho_m$ is the energy density. 
An appropriate distribution of electric charge was chosen, cf. Eq.~\eqref{eq:cargain}. 
The surface layer (a thin shell) is a timelike (lightlike in a special
limit) surface and its matter content is assumed to be a perfect fluid such that 
the superficial energy density 
($\sigma$) and the intrinsic pressure ($\mathcal{P}$) are related by a barotropic
equation of state $\mathcal{P}=\omega\sigma$, $\omega$ being a constant parameter. 
In the lightlike non-singular case the matter quantities on the shell vanish.

The model presents two solutions for the thin shell mass,  and we analyzed several different 
types of configurations for four specific choices of the parameter $\omega$ in each case.  
For the range of parameters of interest, the solutions we found may represent electrically 
charged regular black holes, quasiblack holes, and very compact regular charged stars, besides other less interesting objects.

The stability analysis of the compact objects presented in this manuscript is an undergoing task, and the results will be published elsewhere.

\section*{Acknowledgments}

This work is partly supported by Funda\c{c}\~ao de Amparo
\`a Pesquisa do Estado de S\~ao Paulo (FAPESP, Brazil), Grant No. 
2015/26858-7. We also thank partial financial support from 
Conselho
Nacional de Desenvolvimento Cient\'\i fico e Tecnol\'ogico (CNPq, Brazil), 
Grant No.~308346/2015-7, and from Coordena\c{c}\~ao de
Aperfei\c{c}oamento do Pessoal de N\'\i vel Superior (CAPES, Brazil),
 Grant No.~88881.064999/2014-01.

\end{document}